\documentclass[twocolumn,journal]{IEEEtran}
\IEEEoverridecommandlockouts
\usepackage[T1]{fontenc}%
\usepackage[cmintegrals]{newtxmath}
\usepackage[mathcal]{euscript}
\usepackage{booktabs}
\usepackage{graphicx}
\graphicspath{{./}}
\DeclareGraphicsExtensions{.pdf,.eps,.jpeg,.png}
\usepackage{epstopdf}
\AppendGraphicsExtensions{.tif}
\usepackage[export]{adjustbox}
\usepackage{color}

\usepackage{grffile}
\newlength\figureheight
\newlength\figurewidth
\newlength\subfloatwidth
\usepackage{amsmath}
\usepackage{cite}
\usepackage{amssymb}
\usepackage{dsfont}

\usepackage{algpseudocode}

\usepackage{algorithm}

\usepackage{fixltx2e}

\usepackage[caption=false,subrefformat=parens,labelformat=parens,font=footnotesize]{subfig}

\DeclareMathOperator{\E}{\mathbb{E}}

\newlength{\singlecolumn}
\newcounter{mytempeqncnt}

\title{Learning Wireless Networks' Topologies Using Asymmetric Granger Causality}
\author{Mihir~Laghate,~\IEEEmembership{Student~Member,~IEEE,} and
        Danijela~Cabric,~\IEEEmembership{Senior~Member,~IEEE}%
\thanks{Mihir Laghate is with Qualcomm Technologies Inc., San Diego, CA 92121 and Danijela Cabric is with the Department of Electrical Engineering, University of California, Los Angeles (UCLA), Los
Angeles CA 90095. This work was done when Mihir Laghate was at UCLA.}
\thanks{E-mail: mvlaghate@ucla.edu and danijela@ee.ucla.edu.}
\thanks{This work has been supported by the National Science Foundation under grant 1527026.}}

\begin{document}
\maketitle
\begin{abstract}

    Sharing spectrum with a communicating incumbent user (IU) network requires avoiding interference to IU receivers.  
    But since receivers are passive when in the receive mode and cannot be detected, the network topology can be used to 
    predict the potential receivers of a currently active transmitter. For this purpose, this paper proposes a method to 
    detect the directed links between IUs of time multiplexing communication networks from their transmission start and 
    end times.  It models the response mechanism of commonly used communication protocols using Granger causality: the 
    probability of an IU starting a transmission after another IU's transmission ends increases if the former is a 
    receiver of the latter. This paper proposes a non-parametric test statistic for detecting such behavior. To help 
    differentiate between a response and the opportunistic access of available spectrum, the same test statistic is used 
    to estimate the response time of each link. The causal structure of the response is studied through a discrete time 
    Markov chain that abstracts the IUs' medium access protocol and focuses on the response time and response 
    probability of 2 IUs.  Through NS-3 simulations, it is shown that the proposed algorithm outperforms existing 
    methods in accurately learning the topologies of infrastructure-based networks and that it can infer the directed 
    data flow in ad hoc networks with finer time resolution than an existing method.

\end{abstract}


\IEEEpeerreviewmaketitle

\section{Introduction}
\label{sec:topology_introduction}

Consider the scenario where a cognitive network is sharing spectrum with one or more incumbent networks. Spectrum
sharing by opportunistic spectrum access requires that the cognitive network avoid interference to the receivers in the
incumbent networks. At present, there is extensive work on the detection of transmitters in the field of spectrum
sensing~\cite{ali_advances_2017} but significantly less on identifying receivers.
In particular, existing methods for coexistence in TV white space~\cite{ieee_80222_2011} and the Spectrum Access System
(SAS) of the Citizen's Broadband Radio Service (CBRS)~\cite{fcc_cbrs_2015} detect transmissions and enforce a protection
region around the transmitters where CRs are not allowed to transmit. The size of this protection region is designed
such that the CRs outside the protection region do not cause harmful interference to any incumbent receivers located on
the border of the incumbent transmitter's service area. On the other hand, if we can identify the potential receivers,
we can enforce a smaller protection region around the receivers rather than the transmitters. 

Identifying the current receivers in a network of communicating IUs is difficult because receivers are inherently 
passive when they are in receiving mode and because it is costly for the cognitive network to decode the packets in the 
incumbent network.  Instead, we can use the IUs' network topology to identify the potential receivers of currently 
active transmitters. This observation motivates us to learn the IUs' network topology, i.e., the directed links between 
IUs. Similar to existing work, we will focus on learning the network topology of time multiplexing incumbent networks.

\begin{figure}
    \centering
    \footnotesize
    \includegraphics{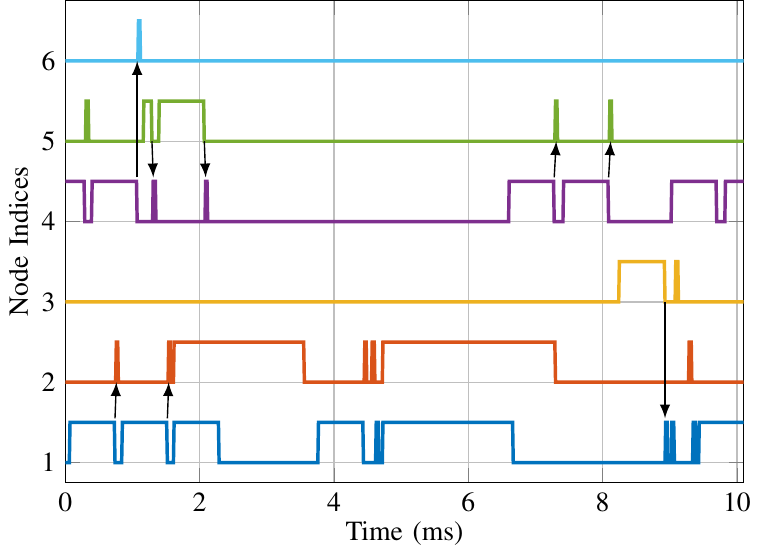}
    \caption{Example activity traces of 802.11n networks from NS-3 simulation for a 6 IU system
    consisting of two infrastructure based networks with APs indexed 1 and 4. Arrows indicate some transmit-response
    pairs.}
    \label{fig:topology_activity}
\end{figure}

\subsection{Existing Work}

In literature, the temporal patterns in IU transmissions have been used to learn network topology. The features used
include transmission start and end times, frame durations, and inter-arrival times. We shall base our work on a similar
input. As has been shown in~\cite{laghate_cooperatively_2017}, we do not need prior knowledge of the IUs' communication
protocols to distinguish IUs and detect the transmission activities of each IU. Hence, we do not require knowledge of
the IUs' communication protocol to obtain these inputs. Note that the location of the IUs is not useful in learning
their network topology because the networks need not be geographically separated. 

One set of existing methods to learn network topology are based on the causal nature of time multiplexing communication
protocols: A transmission by an IU is likely to cause a response, say in the form of an acknowledgement, by its intended
receiver after a fixed but unknown delay.  Examples of such networks include the 802.11 family of protocols, Bluetooth,
and TDD-LTE. Fig.~\ref{fig:topology_activity} shows an example of the transmit-response pairs in activity sequences of
802.11n networks simulated in NS-3. Using this observation, \cite{tilghman_inferring_2013,moore_analysis_2016} model the
responses as a causal structure in the IUs' transmission activity time series. They propose methods to identify pairs of
communicating IUs by detecting IUs that have a higher probability of transmitting after another. This paper is also
based on the same idea.

Another approach to this problem recognizes that the directed links in an ad hoc network correspond to the end-to-end
routes and learns those routes directly~\cite{kokalj-filipovic_learning_2016}. The authors propose an algorithm that
clusters IUs that have similar distributions of packet lengths and inter-arrival times to infer that these IUs
participate in the same data flow. Since they do not infer the ordering of the IUs in the route, we need additional
information, e.g., locations, to identify potential receivers of each IU along the route. Their method also requires the
network to have the same data rates on all links. In addition, in infrastructure-based networks, receiving IUs typically
do not repeat the transmission and are unlikely to have the same packet lengths as the transmitting IU.

\subsection{Challenges and Contributions}

The primary challenge with using causality to detect responses is that if two IUs are not communicating with each other
but are located close enough to interfere with each other, then their collision avoidance mechanisms will ensure that
their transmission activities are not independent. This implies that, unlike typical Granger causality literature, the
causality between two IUs is non-zero even if they are not communicating with each other.

The second challenge is the time varying nature of network topology: the inferences of our algorithm are only useful if 
they are valid for some period of time. We will show that the proposed algorithm learns network topology in the order of 
100ms. Since network topology does not change that frequently, cognitive radios can use the learned topology for making 
inferences for some period of time in the future.

Thirdly, it can be argued that inferences on the entire network should be made simultaneously through a multivariate
method (such as graphical LASSO) since there will be multiple causal relationships in the incumbent network. Instead, we
will use a common feature of communication protocols to argue that testing each pair of IUs independently is sufficient
for learning the network topology rather than making an inference for the entire network simultaneously. Note
that~\cite{tilghman_inferring_2013} also tests pairs of IUs independently.

To help ensure that we detect responses, we choose to detect the causality from the transmission end times of
transmitter IU to the transmission start times of the receiver IU.  Hence, we propose a binary hypothesis test for each
ordered pair of IUs based on the asymmetric form of Granger causality as compared to the symmetric form used
in~\cite{tilghman_inferring_2013,moore_analysis_2016}. For this test, we propose a non-parametric test statistic that
does not assume a linear model between the transmission start and end times of the pair of IUs being tested.

Since such a test may confuse frequent opportunistic spectrum access as responses, we study the response time in
existing time multiplexing communication protocols to note that the response time is shorter than the idle time and is
required to be roughly constant. Based on these observations and verification from NS-3 simulations, we propose an
algorithm to estimate the response time from the same test statistic mentioned above. We call the combination of the
test statistic and this estimation algorithm as the Asymmetric Transfer Entropy to Learn Network Topology (ATELNeT)
algorithm.

To study the effects of dependent transmission activities, we propose a Markov chain model that abstracts the IUs'
medium access protocol and focuses on the response time and retransmission times of a pair of IUs. From this model, we
verify that the response time will be estimated correctly and show that both false alarm as well as detection
probability increase as the frames become shorter and more frequent.

The rest of this paper is organized as follows. Section~\ref{sec:topology_sys_model} describes our system model and
assumptions.  Section~\ref{sec:topology_granger} defines Granger causality, relates different tests from literature to
existing work on topology learning, and proposes a non-parametric test statistic for asymmetric Granger causality.
Section~\ref{sec:topology_proposed} motivates and proposes our ATELNeT algorithm. The Markov chain model for 2 IUs is
discussed in Section~\ref{sec:topology_analysis}. NS-3 simulations of infrastructure-based and ad hoc networks are
presented in Section~\ref{sec:topology_simulations}. The paper is concluded in Section~\ref{sec:topology_conclusion}.

\section{System Model}
\label{sec:topology_sys_model}

We consider a scenario with $M$ IUs indexed as $1,\ldots,M$. We assume that these IUs have been identified by a system
such as that proposed in~\cite{laghate_cooperatively_2017}, by decoding physical layer headers, localization, or
fingerprinting their radios.  In addition, we assume that this system provides a time sequence of the activity of each
IU, sampled periodically at $T_s$ intervals, such that the activity of the $m$'th IU is denoted by $a_m[t] = 1$ if it is
transmitting at time $tT_s$ and 0 otherwise. For the purpose of this work, we shall assume that these sequences are
error-free and that each sequence is of length $N$ samples.

We denote the pair-wise links between these IUs by a binary matrix $L \in \{0,1\}^{M \times M}$ such that $L_{i,j} = 1$
if there is a link from IU $i$ to IU $j$.

For a time series $X[t]$ and positive integer $\tau$, we use $X^{(\tau)}[t]$ to denote $X[t-1], \ldots, X[t-\tau]$. We
use $\mathds{1}_{\{\,\cdot\,\}}$ to denote an indicator variable. For developing our algorithm,
we will denote the sampled time series of transmission start time indicators of the $i$'th IU by $S_i[t] \triangleq
\mathds{1}_{\left\{a_i[t] = 1, a_i[t - 1] = 0\right\}}$. We denote the sampled indicator time series of the end of
transmissions of the $i$'th IU by $E_i[t] \triangleq \mathds{1}_{\left\{a_i[t + 1] = 0, a_i[t] = 1\right\}}$.

\subsection{Assumptions about IUs' Medium Access Protocol}
\label{sec:topology_assumptions}

We define a transmit-receive link by the receiver's act of responding to a signal by the transmitter. Consider a link
from IU $i$ to IU $j$. Assuming a half duplex communication link, the receiver $j$ has to wait for the transmitter $i$
to finish transmitting before $j$ can respond. Thus, the end of a transmission by IU $i$ causes the start of a
transmission by IU $j$ after a short time interval. We call this time interval as the response time or causal lag and
denote it by $\tau_{i,j}$. It consists of both processing times as well as hardware delays such as the Rx-to-Tx
turnaround time. For example, this is defined in the IEEE 802.11 standard~\cite[\S 10.3.7]{ieee_80211_2016} as the Short
Inter-Frame Space (SIFS) which is computed as
\begin{align*}
    \text{aSIFSTime} = & \text{ aRxPHYDelay} + \text{aMACProcessingDelay} \notag \\
                       & + \text{aRxTxTurnaroundTime}.
\end{align*}
Similarly, LTE-TDD has a fixed length guard period between downlink and uplink pilots as part of a special
subframe~\cite[\S 4.2]{etsi_lte_2017}. Bluetooth~\cite{bluetoothsig_bluetooth_2016} has a fixed interframe space
\texttt{T\_IFS} between every Master to Slave and Slave to Master transmission. Finally, each of these standards define
a maximum allowed variation in the response time. It is to be expected that such a response time will feature in future
time division multiplexing communication protocols. Hence, we incorporate the response time in our system model and say
that there is link $i \rightarrow j$ if $E_i[t] = 1$ causes $S_j[t + \tau_{i,j}] = 1$ where $\tau_{i,j}$ is the response
time of that link.  Note that we do not mean to assume that $j$ transmits a response to every transmission by $i$, but
if $j$ does respond to $i$'s transmission, then it will do so $\tau_{i,j}$ time after $i$'s transmission ends. We also
assume that the variation in response times on a link is significantly smaller than $T_s$.

Our second assumption is that the time interval between two successive transmissions by an IU is longer than the
response time for any link originating from that IU.  Mathematically, 
\begin{equation*}
    P(a_i[n + t] = 1 | a_i[n - 1] = 1, a_i[n] = 0) = 0
\end{equation*}
if $t < \max_{j \in \{1,\ldots,M\}}\{\tau_{i, j} : L_{i,j} = 1\}$.
This assumption is based on the fact that communication protocols are designed to avoid interference to the response.
For example, the 802.11 standard specifically sets the NAV field to include the acknowledgement frame's
duration~\cite{ieee_80211_2016}.

Finally, we assume that we know an upper bound $\tau_{\max}$ on $\tau_{i,j}$ for all $i, j \in \{1,\ldots,M\}$ such that
$i \rightarrow j$.

\section{Granger Causality}
\label{sec:topology_granger}

Similar to existing work, we are going to base our work on the idea of Granger causality. In this section, we briefly
introduce the concept, theoretical models, and tests used in existing literature. We relate these methods to the
existing works on learning network topology and then propose a new test statistic for testing \emph{asymmetric} Granger
causality.

The concept of Granger causality was first proposed for macroeconomic analysis~\cite{granger_investigating_1969}.  As
defined in~\cite{granger_investigating_1969}, a time series $X[t]$ is said to Granger cause a time series $Y[t]$ if the
minimum mean square error (MMSE) estimator of $Y[t]$ given $X^{(\infty)}[t]$ and $Y^{(\infty)}[t]$ has lower error than
the MMSE estimator of $Y[t]$ given only $Y^{(\infty)}[t]$. Since the MMSE estimator of $Y[t]$ given $X^{(\infty)}[t]$ 
and
$Y^{(\infty)}[t]$ is $\E[Y[t]|X^{(\infty)}[t],Y^{(\infty)}[t]]$~\cite{sayed_adaptive_2011}, we can say that $X$ Granger
causes $Y$ if $Y[t]$ is not conditionally independent of $X^{(\infty)}[t]$ given $Y^{(\infty)}[t]$. The causal lag
defined in~\cite{granger_investigating_1969} corresponds to the sampled response time $\tau_{\cdot, \cdot}$ in our
system model.  Since the causal lag is unknown, it is usually estimated from the input data by either the Akaike
Information Criterion (AIC) or Bayesian Information Criterion (BIC)~\cite{seth_matlab_2010}. Thus, the problem of
testing Granger causality between a pair of time series $X$ and $Y$ is a binary hypothesis problem where the null
hypothesis ($\mathcal{H}_0$) is that $X$ does not Granger cause $Y$.

\subsection{Models and Tests}

Most methods in literature for testing Granger causality are based on regression but are modified with domain knowledge
specific to the problem under consideration~\cite{seth_matlab_2010}. Specifically, a causal link from $X[t]$ to $Y[t]$
is modeled as a vector autoregressive model such as~\cite{granger_investigating_1969,seth_matlab_2010}
\begin{equation}
    Y[t] = \sum_{\tau = 1}^{\tau_{\max}} \alpha_{1,\tau} X[t - \tau] + \sum_{\tau = 1}^{\tau_{\max}} \beta_{1,\tau} Y[t-\tau] +
    \xi_1[t]
    \label{eq:topology_lin_alt}
\end{equation}
where $\alpha_{1,\tau}$ and $\beta_{1,\tau}$ are the parameters of the regression and $\xi_1[t]$ is the residual noise.
Similarly, the null hypothesis is modeled as
\begin{equation}
    Y[t] = \sum_{\tau = 1}^{\tau_{\max}} \beta_{0,\tau} Y[t-\tau] + \xi_0[t]
    \label{eq:topology_lin_null}
\end{equation}
with $\beta_{0,\tau}$ being the regression parameters and $\xi_0[t]$ the residual noise. Therefore, the null hypothesis
is modeled by the coefficients $\alpha_{1,\tau}$ in this model being zero. 

In~\cite{tilghman_inferring_2013}, the authors use this linear model for testing causality between every pair of IUs on
disjoint observation windows. Their hard fusion algorithm then fuses the topologies learned in each window using the
majority rule. For individual windows, the hard fusion algorithm proposed in~\cite{tilghman_inferring_2013} uses the
test statistic
\begin{align}
    g_{i \rightarrow j} & \triangleq \left(\frac{\sum_{t = 1}^{N - \tau} |\hat{\xi}_0[t]|^2 - \sum_{t = 1}^{N - \tau} |\hat{\xi}_1[t]|^2}{\sum_{t = 1}^{N - \tau}
    |\hat{\xi}_1[t]|^2} \right) \frac{N - 3\tau - 1}{\tau}
    \label{eq:topology_tilghman_hard} \\
    & \sim F(\tau, N - 3\tau - 1) \label{eq:topology_tilghman_hard_distrib}
\end{align}
where $\hat{\xi}_1[t]$ and $\hat{\xi}_0[t]$ are the estimates of $\xi_1[t]$ and $\xi_0[t]$ respectively and the authors
assume that both $\xi_0[t]$ and $\xi_1[t]$ have zero-mean Gaussian distributions and equal variances. The threshold is
chosen using (\ref{eq:topology_tilghman_hard_distrib}) to satisfy a given false alarm probability. In
their soft fusion algorithm, the authors compute the average causality magnitude~\cite{geweke_measurement_1982} of a
link
\begin{equation}
    F_{i \rightarrow j} = \ln \left( \frac{\E\left[|\xi_1[t]|^2\right]}{\E\left[|\xi_0[t]|^2\right]} \right).
    \label{eq:topology_causality_magnitude}
\end{equation}
for each window. The network topology is inferred as those links with average causality magnitude greater than the
average of all pairs of IUs.

Transmission start times have been considered as continuous time point processes and modeled as multivariate Hawkes
processes in~\cite{moore_analysis_2016}. In the Hawkes process model, an event in one time series increases the
probability of an event occurring on other time series for a short period of time. The support of these impact functions
were recently shown to be equivalent to Granger causality~\cite{eichler_graphical_2017}. There is a rich body of
literature learning Hawkes point processes and might be useful for learning Granger causality in multivariate systems as
done in~\cite{xu_learning_2016}. With respect to our system model, we are using sampled activity sequences as input and
not point processes.

Finally, an information theoretic non-parametric test for Granger causality has been proposed
in~\cite{schreiber_measuring_2000}. The author of~\cite{schreiber_measuring_2000} proposed the conditional mutual
information of $Y[t]$ and $X^{(\tau)}[t]$ given $Y^{(\tau)}[t]$:
\begin{align}
    T_{X \rightarrow Y} \triangleq \sum & \Bigg[ P\left(Y[t], X^{(\tau)}[t], Y^{(\tau)}[t]\right) \notag \\
        & \quad \left. \times \log \frac{P\left(Y[t] |
    X^{(\tau)}[t], Y^{(\tau)}[t]\right)}{P\left(Y[t] | Y^{(\tau)}[t]\right)} \right]
    \label{eq:topology_te_def}
\end{align}
as a test statistic called transfer entropy for testing whether $X[t]$ Granger causes $Y[t]$. In
(\ref{eq:topology_te_def}), the summation is over all possible values of $X^{(\tau)}[t]$,$Y^{(\tau)}[t]$, and $Y[t]$.  
The author proposes a binary hypothesis test with $T_{X \rightarrow Y} = 0$ as the null hypothesis for no causality and 
$T_{X \rightarrow Y} > 0$ as the alternate hypothesis denoting a causal relationship.
We shall base our proposed method on a similar test statistic described below.

\subsection{Asymmetric Granger Causality}
\label{sec:topology_granger_ate}

In this work, we are going to use an asymmetric form of Granger causality loosely based on that proposed
in~\cite{hatemi-j_asymmetric_2012}\footnote{\cite{hatemi-j_asymmetric_2012} separates events occurring on the time 
series as we do, but tests for conventional Granger causality between the generated time series. We propose 
incorporating a third time series to model a common cause for both hypotheses}.  Consider three time series $X[t]$, 
$Y_1[t]$, and $Y_2[t]$.  We will say that $X[t]$
asymmetrically Granger causes $Y_2[t]$ given $Y_1[t]$ if $Y_2[t]$ and $X^{(\infty)}[t]$ are not conditionally
independent given $Y_1^{(\infty)}[t]$. Note the absence of $Y^{(\infty)}_2[t]$ in this definition. Thus, we can think of
$X[t]$ and $Y_1[t]$ as events representing potential causes while $Y_2[t]$ denotes the events representing effects.

Here, $Y_1[t]$ and $Y_2[t]$ typically describe different events on an underlying common
sequence while $X[t]$ describes events on a second time sequence. For example, in our proposed algorithm, $Y_1[t]$ and
$Y_2[t]$ will correspond to the end and start transmission events respectively of one IU while $X[t]$ will denote the end
transmission events of another IU.

Similar to Granger causality, we use a finite history $\tau$ while testing asymmetric Granger causality. We
propose an asymmetric transfer entropy as our test statistic
\begin{align}
A_{X \rightarrow Y_2|Y_1}(\tau) \triangleq \sum & \Bigg[ P\left(Y_2[t], X^{(\tau)}[t], Y_1^{(\tau)}[t]\right) \notag \\
& \quad \left. \times \log \frac{P\left(Y_2[t] | X^{(\tau)}[t], Y_1^{(\tau)}[t]\right)}{P\left(Y_2[t] | 
    Y_1^{(\tau)}[t]\right)} \right]
    \label{eq:topology_ate_def}
\end{align}
where the summation is over all values taken by $X^{(\tau)}[t]$, $Y_1^{(\tau)}[t]$, and $Y_2[t]$. The test statistic 
shall compare between the same hypotheses as described above for~\cite{schreiber_measuring_2000}. The motivation and use 
of this test statistic for learning network topology is described in the next section.

\section{Proposed Method: Asymmetric Transfer Entropy to Learn Network Topology (ATELNeT)}
\label{sec:topology_proposed}

In this section, we will be proposing a test for detecting causal links between ordered pairs of IUs at a time. In
general, it is true that there may be a performance benefit in detecting all links simultaneously through a multivariate
test.  However, as mentioned earlier in Section~\ref{sec:topology_assumptions}, communicating protocols are usually
designed to avoid interfering with the response of the receiver. Hence, by focusing on the response time of each link,
we expect that other IUs' transmissions will not interfere with the testing of a given link. Hence, we also propose a
method for estimating the response time from the test statistic. We will also be proposing an additional linear
regression based test similar to that of~\cite{tilghman_inferring_2013} for comparing the performance gain from the
choice of nonlinear model against the linear model.

\subsection{Proposed Test Statistic for Transmit-Receive Pair}

\begin{figure}
    \centering
    \includegraphics{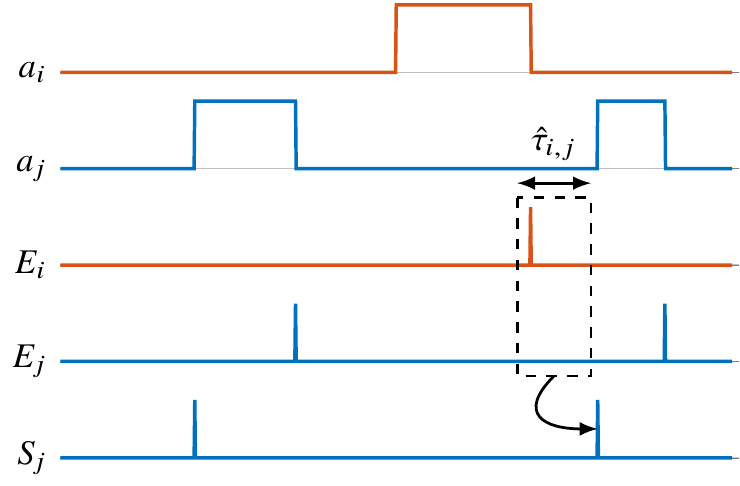}
    \vspace*{-1em}
    \caption{Example inputs to our proposed algorithm for testing link $i \rightarrow j$. The algorithm tests the
             causality from the input values in the dashed rectangle to the transmission start indicator.  The response
             time $\tau_{i,j}$ needs to be estimated.}
    \label{fig:topology_ate_inputs}
    \vspace*{-1em}
\end{figure}

As mentioned above, we propose testing every ordered pair $i$, $j$ of IUs independently. Similar 
to~\cite{schreiber_measuring_2000}, we consider two hypotheses: the
null hypothesis $\mathcal{H}_0(i,j)$ represents $i \not\rightarrow j$ and the alternate hypothesis $\mathcal{H}_1(i,j)$
represents $i \rightarrow j$. To specifically detect the responses transmitted by IU $j$ to transmissions of IU $i$, we
test whether $E_i[t]$ asymmetrically Granger causes $S_j[t]$ given $E_j[t]$. Fig.~\ref{fig:topology_ate_inputs} shows
the inputs that we use for testing this asymmetric causality.  Since there is no reason to assume a linear model for
$E_i[t]$, $E_j[t]$, and $S_j[t]$, we use the non-parametric asymmetric transfer entropy (\ref{eq:topology_ate_def}) that
we proposed in Section~\ref{sec:topology_granger_ate}:
\begin{equation}
    \hat{A}_{E_i \rightarrow S_j|E_j}(N, \hat{\tau}_{i,j}) \mathop{\gtrless}_{\mathcal{H}_0(i,j)}^{\mathcal{H}_1(i,j)}
    \lambda(\hat{\tau}_{i,j})
\end{equation}
where $\hat{A}_{E_i \rightarrow S_j|E_j}(N, \hat{\tau}_{i,j})$ is the empirical estimate of $A_{E_i \rightarrow
S_j|E_j}(\hat{\tau}_{i,j})$ computed from $N$ activity samples with a response time of $\hat{\tau}_{i,j}$,
$\lambda(\hat{\tau}_{i,j})$ is a threshold that depends on the response time. The response time $\hat{\tau}_{i,j}$ will
need to be estimated from the data and the threshold $\lambda(\,\cdot\,)$ will be computed from the distribution of
$\hat{A}_{E_i \rightarrow S_j|E_j}$ as described next.

We derive the asymptotic distribution of $\hat{A}_{E_i \rightarrow S_j|E_j}$ in 
Appendix~\ref{appendix:topology_ate_distribution}.
Our derivation is similar to that of the asymptotic distribution of $\hat{T}_{X \rightarrow Y}$ 
in~\cite{barnett_transfer_2012}.  We
show that for large $N$ and a given causal lag $\tau$, $2(N - \tau)\hat{A}_{E_i \rightarrow S_j|E_j}(N, \tau)$ has
distribution $\chi^2_{d_{i,j}}$ under the null hypothesis $\mathcal{H}_0(i,j)$ and distribution
$\chi^2_{d_{i,j}}\left(2(N-\tau) A_{E_i \rightarrow S_j|E_j}(\tau) \right)$ under the alternate hypothesis
$\mathcal{H}_1(i,j)$ where $\chi^2_d$ is a central $\chi^2$ distribution with $d$ degrees of freedom and $\chi^2_d(c)$
is a non-central $\chi^2$ distribution with $d$ degrees of freedom and non-centrality parameter $c$.

We choose the threshold $\lambda(\tau)$ such that the probability of false alarm is asymptotically bounded above by a 
parameter $P_{\text{FA}}$:
\begin{equation}
    \lambda(\tau) \triangleq \chi^{-2}_{d_{i,j}}(1 - P_\text{FA})
    \label{eq:topology_ate_threshold}
\end{equation}
where $\chi^{-2}_d(\,\cdot\,)$ is the inverse cumulative distribution function of a central $\chi^2$ random variable
with $d$ degrees of freedom.

Note that we have assumed that $A_{E_i \rightarrow S_j|E_j}(\tau) = 0$ under the null hypothesis. Though this is a
common hypothesis in the Granger causality literature as well as~\cite{tilghman_inferring_2013,moore_analysis_2016}, we
will show below that $A_{E_i \rightarrow S_j|E_j}(\tau) > 0$ even if $\mathcal{H}_0(i, j)$. Since this value is unknown 
a priori, we choose $A_{E_i \rightarrow S_j|E_j}(\tau) = 0$ to define the null hypothesis.  Fortunately, both the model 
analyzed in Section~\ref{sec:topology_analysis} and simulations in Section~\ref{sec:topology_simulations} show that 
$A_{E_i \rightarrow S_j|E_j}(\tau)$ is sufficiently small that we can still use this typical null hypothesis of $A_{E_i 
\rightarrow S_j|E_j}(\tau) = 0$.

Next, we need to estimate the degrees of freedom $d_{i,j}$. As per the derivation in
Appendix~\ref{appendix:topology_ate_distribution}, $d_{i,j}$ is the difference in the number of parameters for models
under the alternate hypothesis $\mathcal{H}_1(i,j)$ and the null hypothesis $\mathcal{H}_0(i,j)$. Since $E_i[t], E_j[t],
S_j[t] \in \{0,1\}$, the number of parameters in either model is simply the number of independent values in the
appropriate joint probability mass function. Next, our assumption that the retransmission time is longer than the
response time means that if $E_j[t - \tau] = 1$ for some $1 \leq \tau \leq \hat{\tau}_{i,j}$ then $S_j[t] = 0$. It
also means that $\sum_{\tau = 1}^{\hat{\tau}_{i,j}} E_j[t - \tau] \leq 1$ and $\sum_{\tau = 1}^{\hat{\tau}_{i,j}} E_i[t
- \tau] \leq 1$. Hence, the number of parameters under null hypothesis is $\hat{\tau}_{i,j} + 1$ and under the alternate
hypothesis is $(\hat{\tau}_{i,j} + 1)^2$. Hence, the degrees of freedom are
\begin{equation}
    d_{i,j} = \hat{\tau}_{i,j}(\hat{\tau}_{i,j} + 1).
    \label{eq:topology_dof}
\end{equation}

\subsection{Proposed Algorithm to Estimate Response Time of a Link}
\label{sec:topology_ate_lag}

\begin{figure}
    \centering
    \footnotesize
    \includegraphics{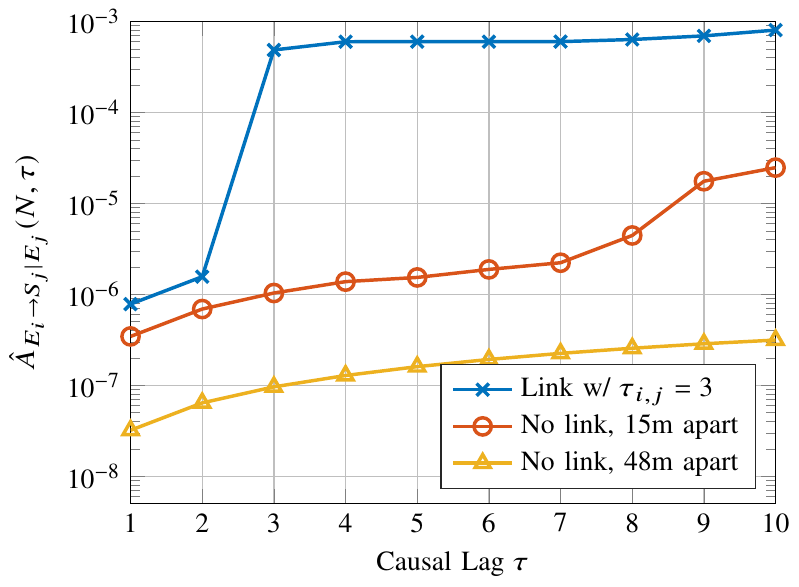}
    \caption{Asymmetric transfer entropy of pairs of IUs in NS-3 simulated 802.11n networks computed with increasing
    causal lags. Parameters: $T_s \ 5\mu$s and observation duration 10s.}
    \label{fig:topology_ate_lag}
\vspace*{-2em}
\end{figure}

Estimating the response time $\hat{\tau}_{i,j}$ depends on its invariance. In particular, if the variance in response
time is significantly smaller than the sampling time of the link, then the response is likely to be received at either
$\tau_{i,j}$ or $\tau_{i,j} + 1$ with high probability. Hence, $A_{E_i \rightarrow S_j|E_j}(\tau_{i,j})$ will be
significantly higher than $A_{E_i \rightarrow S_j|E_j}(\tau_{i,j} - 1)$. This intuition will be revisited analytically
in Section~\ref{sec:topology_analysis}. From NS-3 data of 3 pairs of 802.11n IUs, Fig.~\ref{fig:topology_ate_lag} shows
that $\hat{A}_{E_i \rightarrow S_j|E_j}(N, \tau)$ increases sharply at $\tau = 3$, i.e., after SIFS of $16\mu$s, if
there is a link but not if the pair do not have a link.  For the pair of IUs that do not have a link but are close to
each other, Fig.~\ref{fig:topology_ate_lag} also shows a slight increase in $\hat{A}_{E_i \rightarrow S_j|E_j}(N, 
\tau)$ at the lags after the minimum retransmission time, i.e., after DIFS of 34$\mu$s.

Based on this discussion, we estimate $\hat{\tau}_{i,j}$ such that
\begin{equation}
    \hat{A}_{E_i \rightarrow S_j|E_j}(N, \hat{\tau}_{i,j}) > \alpha \hat{A}_{E_i \rightarrow S_j|E_j}(N,
    \hat{\tau}_{i,j} - 1)
\end{equation}
for a parameter $\alpha > 1$. Pseudocode for estimating $\hat{\tau}_{i,j}$ is provided in
Algorithm~\ref{algo:topology_ate_lag}. Starting from the maximum response time $\tau_{\max}$, we keep reducing the
estimated response time $\hat{\tau}_{i,j}$ until the asymmetric transfer entropy reduces by the factor $\alpha$. For the
alternate hypothesis, this finds the correct response time as will be seen from the simulations. On the other hand, the
test statistic does not change rapidly under the null hypothesis and the algorithm stops at $\tau = 1$.  As we will
discuss in Section~\ref{sec:topology_analysis}, under the null hypothesis, the asymmetric transfer entropy is minimum at
$\tau = 1$. Hence, this estimate of the response time ensures the best match with the null hypothesis assumption that
$A_{E_i \rightarrow S_j|E_j}(\tau_{i,j}) = 0$.

\begin{algorithm}[t]
    \caption{Estimating response time for link $i \rightarrow j$}
\label{algo:topology_ate_lag}
\begin{algorithmic}[1]
    \Require $\tau_{\max} \in \mathbb{N}$, $\alpha \in \mathbb{R}_+$, $\hat{P}\left(E_i^{(\tau_{\max})}[t], E_j^{(\tau_{\max})}[t], S_j[t]\right)$
    \State $\tau \gets \tau_{\max}$
    \State {$r \gets \hat{A}_{E_i\rightarrow S_j|E_j}(N, \tau - 1) / \hat{A}_{E_i \rightarrow S_j|E_j}(N, \tau)$}
    \While {$\tau > 1$ and $r > \alpha$}
        \State $\tau \gets \tau - 1$
        \State {$r \gets \hat{A}_{E_i\rightarrow S_j|E_j}(N, \tau - 1) / \hat{A}_{E_i \rightarrow S_j|E_j}(N, \tau)$}
    \EndWhile
    \State $\hat{\tau}_{i,j} \gets \tau$
    \Ensure $\hat{\tau}_{i,j}$
\end{algorithmic}
\end{algorithm}

\subsection{Linear Test for Asymmetric Causality}
\label{sec:topology_proposed_linear}

The ATELNeT algorithm proposed above is different from the existing work in two aspects: the choice of testing the
asymmetric rather than symmetric Granger causality and the transfer entropy test statistic as opposed to the regression
based approaches. In order to understand the performance gain of the two aspects separately, we propose a second test
statistic based on linear regression.

Similar to the approach of~\cite{tilghman_inferring_2013} and with notation analogous to that of
(\ref{eq:topology_lin_alt})-(\ref{eq:topology_lin_null}), we model the null hypothesis $\mathcal{H}_0(i,j)$ as
\begin{equation}
    S_j[t] = \sum_{\tau = 1}^{\tau_{i,j}} \beta_{0,\tau} E_j[t-\tau] + \xi_0[t].
\end{equation}
and the alternate hypothesis $\mathcal{H}_1(i,j)$ as
\begin{equation}
    S_j[t] = \sum_{\tau = 1}^{\tau_{i,j}} \alpha_{1,\tau} E_i[t - \tau] + \sum_{\tau = 1}^{\tau_{i,j}} \beta_{1,\tau}
    E_j[t-\tau] + \xi_1[t].
\end{equation}
We use the test statistic of (\ref{eq:topology_tilghman_hard}) and threshold based on
(\ref{eq:topology_tilghman_hard_distrib}) as per the false alarm constraint $P_{\text{FA}}$: $F^{-1}_{\tau_{i,j}, N -
3\tau_{i,j} - 1}(1 - P_{\text{FA}})$ where $F^{-1}_{a,b}$ is the inverse cumulative $F$ distribution with parameters $a$
and $b$. Unlike~\cite{tilghman_inferring_2013}, we do not split the input into windows.

Note that this approach requires separate estimation of the response time $\tau_{i,j}$.

\section{Analysis of Asymmetric Transfer Entropy in Shared Channel}
\label{sec:topology_analysis}

\begin{figure*}
    \centering
    \includegraphics{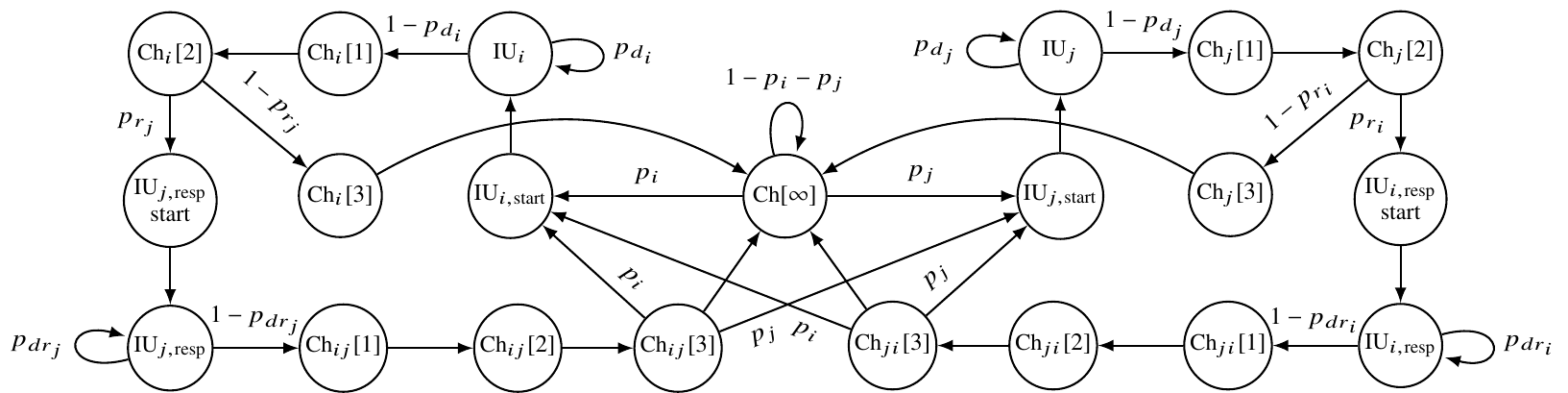}
    \vspace*{-0.6em}
    \caption{Markov chain model for two users sharing a wireless channel under the assumption of no collisions.}
    \label{fig:topology_mc_shared_channel}
    \vspace*{-1.5em}
\end{figure*}

In this section, we propose a Markov chain model of 2 IUs $i$ and $j$ for studying the behavior of $A_{E_i \rightarrow
S_j|E_j}$ as we vary the causal lag $\tau$, the frame lengths, inter-arrival times, and the probability of one IU
responding to another. 
Ideally, this requires modeling the MAC protocol of the incumbent network. But to avoid restricting our study to a
single MAC protocol and for analytical tractability, we re-use the observation in Section~\ref{sec:topology_assumptions}
that IUs will not interfere with each others' responses to construct a simplified discrete time Markov chain model of
only 2 IUs. The Markov chain's state transitions occur every $T_s$ seconds. We make the following assumptions:
\begin{enumerate}
    \item No collisions between IUs $i$ and $j$
    \item Response time $\tau_{ij} = 3 = \tau_{ji}$ for each link
    \item Transmissions and responses are at least 2 samples long
    \item Packet sources have stationary distributions for arrival times and packet lengths
    \item (Frame length - 1) for IU $i$ is a geometric random variable with parameter $1 - p_{di}$
    \item No retransmissions within $\tau_{ij}$ or $\tau_{ji}$.
    \item A response does not cause the source to send a response.
\end{enumerate}
Note that assumption 5 is justified in the system model described in Section~\ref{sec:topology_sys_model}.

We will describe the model from the point of view of IU $i$ transmitting a new frame and IU $j$
responding (or not). The model is symmetric in $i$ and $j$. The states of the model are chosen such that in each state,
the indicator variables $E_i[t]$, $E_j[t]$, $S_i[t]$, and $S_j[t]$ take exactly one value each.

The model begins in the central state $\text{Ch}[\infty]$ that represents the unoccupied channel with no active
transmissions or responses. From this state, either IU may start transmitting a new frame with probability $p_i$ or
$p_j$ respectively. These parameters are related to the frame duration $T_{i,\text{frame}}$ and idle time
$T_{i,\text{idle}}$ (in seconds) of the IUs as follows.
\begin{equation}
    p_{i} = \frac{T_s}{\E[T_{i,\text{frame}}] + \E[T_{i,\text{idle}}]} \text{, }
    p_{j} = \frac{T_s}{\E[T_{j,\text{frame}}] + \E[T_{j,\text{idle}}]} \label{eq:topology_pi_def}
\end{equation}

Without loss of generality, assume IU $i$ starts transmitting. The continued transmission of IU $i$ is modeled by the
state $\text{IU}_{i}$ which has a self loop of probability $p_{d_i}$ to model the frame length as 1 + $G$ where $G$ is a
geometric random variable. The parameter $p_{d_i}$ is related to the expected frame duration (in seconds) by
\begin{equation}
    p_{d_i} = 1 - \frac{T_s}{\E[T_{i,\text{frame}}] - T_s}. \label{eq:topology_pdi_def}
\end{equation}
When the frame ends, the channel is modeled to be vacant for two time samples in states $\text{Ch}_i[1]$ and
$\text{Ch}_i[2]$. Since we have assumed that the response time is 3 samples for both links, the model can transition
from state $\text{Ch}_i[2]$ to a response state $\text{IU}_{j,\text{resp}}~\text{start}$ with a response probability
$p_{r_j}$ or continue to a state of unoccupied channel $\text{Ch}_i[3]$ with probability $1 - p_{r_j}$. If the latter
transition occurs, then the system will simply transition back to the initial unoccupied channel state
$\text{Ch}[\infty]$.

Instead, if IU $j$ begins transmitting a response, then we model the response length $T_{j,\text{response}}$ by the
state $\text{IU}_{j, \text{resp}}$ that has a self loop of probability $p_{dr_j}$:
\begin{equation}
    p_{dr_j} = 1 - \frac{T_s}{\E[T_{j,\text{response}}] - T_s}.
\end{equation}
When the response ends, the channel is modeled as being unoccupied for at least two more time instants because of our
fifth and sixth assumption above. From state $\text{Ch}_{ij}[3]$, either IU may start transmitting with the original
probabilities $p_i$ and $p_j$. If neither begins transmitting, then the channel is unoccupied and the system returns to
the state $\text{Ch}[\infty]$.

The model for IU $j$ transmitting is identical.

It should be noted that the model in Fig.~\ref{fig:topology_mc_shared_channel} defines a scenario where two users
sharing the same channel always detect each other's transmissions and never collide. Hence, the causality in this model
is slightly higher than a more practical scenario.

For the purposes of the following discussion, we consider the null hypothesis to be $\mathcal{H}_0(i,j)$ as defined
earlier. Since the model is symmetric in $i$ and $j$, without loss of generality, we consider only the alternate
hypothesis of $\mathcal{H}_1(i,j)$. Unless mentioned otherwise, we set $p_{r_i} = 0$, i.e., $i$ never responds to $j$'s
transmissions.

\subsection{Computing Asymmetric Transfer Entropy from the Model}
\label{sec:topology_th_ate_comp}

\newcommand{\pstar}{p^*}
\begin{table}
    \centering
    \caption{Values of $E_i^{(3)}[t]$, $E_j^{(3)}[t]$, $S_i[t]$, and $S_j[t]$ for each state and notation for steady state probability}
    \label{tab:topology_state_timeseries}
    \begin{tabular}{l | *{3}{c} | *{3}{c} |c|c|l}
        \toprule
        State & \multicolumn{3}{c|}{$E_i^{(3)}[t]$} & \multicolumn{3}{c|}{$E_j^{(3)}[t]$} & $S_i[t]$ & $S_j[t]$ & Notation\\
        \midrule
        $\text{Ch}[\infty]$        & 0 & 0 & 0 & 0 & 0 & 0 & 0 & 0 & $\pstar_{ch[\infty]}$ \\
        IU$_{i,\text{start}}$      & 0 & 0 & 0 & 0 & 0 & 0 & 1 & 0 & $\pstar_{is}$ \\
        IU$_{i}$                   & 0 & 0 & 0 & 0 & 0 & 0 & 0 & 0 & $\pstar_{i}$ \\
        Ch$_{i}[1]$                & 1 & 0 & 0 & 0 & 0 & 0 & 0 & 0 & $\pstar_{ch_i[1]}$ \\
        Ch$_{i}[2]$                & 0 & 1 & 0 & 0 & 0 & 0 & 0 & 0 & $\pstar_{ch_i[2]}$ \\
        Ch$_{i}[3]$                & 0 & 0 & 1 & 0 & 0 & 0 & 0 & 0 & $\pstar_{ch_i[3]}$ \\
        IU$_{j,\text{resp}}$ start & 0 & 0 & 1 & 0 & 0 & 0 & 0 & 1 & $\pstar_{jr_s}$ \\
        IU$_{j,\text{resp}}$       & 0 & 0 & 0 & 0 & 0 & 0 & 0 & 0 & $\pstar_{jr}$ \\
        Ch$_{ij}[1]$               & 0 & 0 & 0 & 1 & 0 & 0 & 0 & 0 & $\pstar_{ch_{ij}[1]}$ \\
        Ch$_{ij}[2]$               & 0 & 0 & 0 & 0 & 1 & 0 & 0 & 0 & $\pstar_{ch_{ij}[2]}$ \\
        Ch$_{ij}[3]$               & 0 & 0 & 0 & 0 & 0 & 1 & 0 & 0 & $\pstar_{ch_{ij}[3]}$ \\
        IU$_{j,\text{start}}$      & 0 & 0 & 0 & 0 & 0 & 0 & 0 & 1 & $\pstar_{js}$ \\
        IU$_{j}$                   & 0 & 0 & 0 & 0 & 0 & 0 & 0 & 0 & $\pstar_{j}$ \\
        Ch$_{j}[1]$                & 0 & 0 & 0 & 1 & 0 & 0 & 0 & 0 & $\pstar_{ch_j[1]}$ \\
        Ch$_{j}[2]$                & 0 & 0 & 0 & 0 & 1 & 0 & 0 & 0 & $\pstar_{ch_j[2]}$ \\
        Ch$_{j}[3]$                & 0 & 0 & 0 & 0 & 0 & 1 & 0 & 0 & $\pstar_{ch_j[3]}$ \\
        IU$_{i,\text{resp}}$ start & 0 & 0 & 0 & 0 & 0 & 1 & 1 & 0 & $\pstar_{jr_s}$ \\
        IU$_{i,\text{resp}}$       & 0 & 0 & 0 & 0 & 0 & 0 & 0 & 0 & $\pstar_{jr}$ \\
        Ch$_{ji}[1]$               & 1 & 0 & 0 & 0 & 0 & 0 & 0 & 0 & $\pstar_{ch_{ji}[1]}$ \\
        Ch$_{ji}[2]$               & 0 & 1 & 0 & 0 & 0 & 0 & 0 & 0 & $\pstar_{ch_{ji}[2]}$ \\
        Ch$_{ji}[3]$               & 0 & 0 & 1 & 0 & 0 & 0 & 0 & 0 & $\pstar_{ch_{ji}[3]}$ \\
        \bottomrule
    \end{tabular}
    \vspace*{-2em}
\end{table}

\begin{table}
    \centering
    \caption{Probability mass function for $E_i^{(3)}[t]$, $E_j^{(3)}[t]$, $S_i[t]$, and $S_j[t]$.}
    \label{tab:topology_state_steady}
    \begin{tabular}{*{3}{c} | *{3}{c} |c|c| p{8em}}
        \toprule
        \multicolumn{3}{c|}{$E_i^{(3)}[t]$} & \multicolumn{3}{c|}{$E_j^{(3)}[t]$} & $S_i[t]$ & $S_j[t]$ & Probability
        \\
        \midrule
        0 & 0 & 0 & 0 & 0 & 0 & 0 & 0 & $\pstar_{ch[\infty]} + \pstar_i + \pstar_{jr} + \pstar_{ir} + \pstar_j$\\
        0 & 0 & 0 & 0 & 0 & 0 & 0 & 1 & $\pstar_{js}$\\
        0 & 0 & 0 & 0 & 0 & 0 & 1 & 0 & $\pstar_{is}$ \\
        0 & 0 & 0 & 0 & 0 & 1 & 0 & 0 & $\pstar_{ch_j[3]} + \pstar_{ch_{ij}[3]}$\\
        0 & 0 & 0 & 0 & 0 & 1 & 1 & 0 & $\pstar_{ir_s}$ \\
        0 & 0 & 0 & 0 & 1 & 0 & 0 & 0 & $\pstar_{ch_j[2]} + \pstar_{ch_{ij}[2]}$ \\
        0 & 0 & 0 & 1 & 0 & 0 & 0 & 0 & $\pstar_{ch_j[1]} + \pstar_{ch_{ij}[1]}$ \\
        0 & 0 & 1 & 0 & 0 & 0 & 0 & 0 & $\pstar_{ch_i[3]} + \pstar_{ch_{ji}[3]}$ \\
        0 & 0 & 1 & 0 & 0 & 0 & 0 & 1 & $\pstar_{jr_s}$\\
        0 & 1 & 0 & 0 & 0 & 0 & 0 & 0 & $\pstar_{ch_i[2]} + \pstar_{ch_{ji}[2]}$ \\
        1 & 0 & 0 & 0 & 0 & 0 & 0 & 0 & $\pstar_{ch_i[1]} + \pstar_{ch_{ji}[1]}$ \\
        \bottomrule
    \end{tabular}
\end{table}

To compute $A_{E_i \rightarrow S_j|E_j}(\tau_{i,j})$ in this model, we first need to compute the joint probability mass
function of $E_i^{(\tau_{i,j})}[t]$, $E_j^{(\tau_{i,j})}[t]$, and $S_j[t]$. As mentioned above, the states of the model
were designed such that each state had exactly one value for these three time series. This correspondence is listed in
Table~\ref{tab:topology_state_timeseries}. Here, $E_i^{(3)}[t]$ is listed in the order $E_i[t-1], E_i[t-2], E_i[t-3]$.
Similarly for $E_j^{(3)}[t]$. Using this correspondence, we can write the joint probability mass function of
$E_i^{(\tau_{i,j})}[t]$, $E_j^{(\tau_{i,j})}[t]$, and $S_j[t]$ in terms of the steady state probabilities of this Markov
chain model. The joint probability mass function is listed in Table~\ref{tab:topology_state_steady}. 

The steady state probabilities have been derived in Appendix~\ref{appendix:topology_steady_state} by solving the
appropriate simultaneous equations.  The asymmetric transfer entropy for various lags can then be expressed in terms of
our model's parameters by using Table~\ref{tab:topology_state_steady} to compute individual terms in the summation of
(\ref{eq:topology_ate_def}). These expressions for causal lags up to 3 are derived in
Appendix~\ref{appendix:topology_ate_derivation}. Unfortunately, the algebraic expressions themselves do not provide much
insight into the behavior of the test statistic due to the large number of terms involved. Therefore, we study the
behavior of the test statistic numerically by varying its various parameters, viz., the probability of starting a new
transmission $p_i$ and $p_j$, the frame lengths through $p_{d_i}$ and $p_{d_j}$, the probability of IUs responding to
each other $p_{r_i}$ and $p_{r_j}$, and the length of the response frames through $p_{dr_i}$ and $p_{dr_j}$. 

\subsection{Effect of increasing causal lag}
\label{sec:topology_th_lag}

\begin{figure}
    \centering
    \footnotesize
    \includegraphics{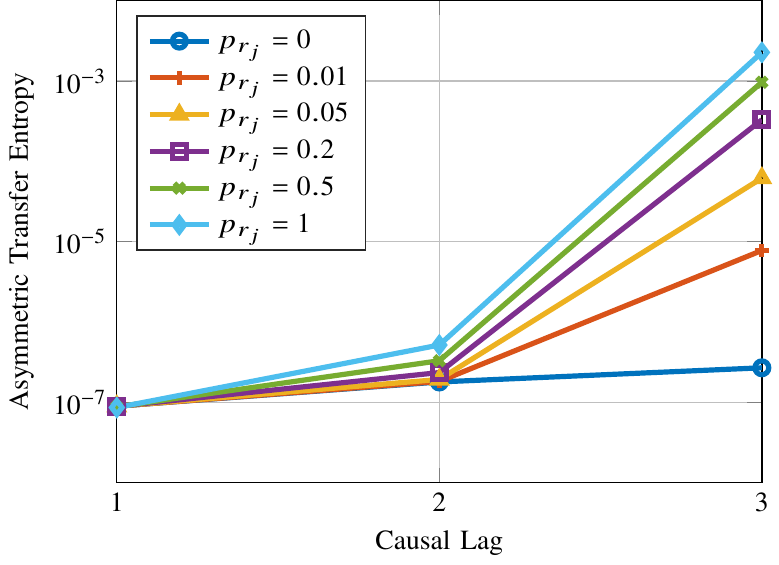}
    \caption{Asymmetric transfer entropy for $i \rightarrow j$ computed from our model for various response
        probabilities and causal lags. System: Expected $T_{\text{on}} + T_{\text{off}} = 10$ms, frame duration 3.33ms,
    response duration 333$\mu$s, $T_s = 5\mu$s.}
    \label{fig:topology_mc_ate_vs_lag}
    
\end{figure}

We begin by studying the increase of asymmetric transfer entropy for the link $i \rightarrow j$ with increasing causal
lag. As Fig.~\ref{fig:topology_mc_ate_vs_lag} shows, $A_{E_i \rightarrow S_j|E_j}(\tau)$ increases with $\tau$ for all
response probabilities $p_{r_j}$, but the rate of increase also increases with $p_{r_j}$. Most importantly, $A_{E_i
\rightarrow S_j|E_j}(\tau)$ increases significantly at $\tau = 3$. This matches our intuition in
Section~\ref{sec:topology_ate_lag} where we expected this sudden increase due to the fact that a response is likely to
occur with high probability at lag 3. We can also use Fig.~\ref{fig:topology_mc_ate_vs_lag} to choose the parameter
$\alpha$ in Algorithm~\ref{algo:topology_ate_lag}. Fig.~\ref{fig:topology_mc_ate_vs_lag} shows that $A_{E_i \rightarrow
S_j|E_j}(3)$ is at least 10 times larger than $A_{E_i \rightarrow S_j|E_j}(2)$ for even $p_{r_j} = 0.01$. Hence, we
suggest $\alpha = 10$ is an appropriate choice for the parameter.

\subsection{Effect of frame durations and idle time}
\label{sec:topology_th_pstart}

\begin{figure}
    \centering
    \footnotesize
    \includegraphics{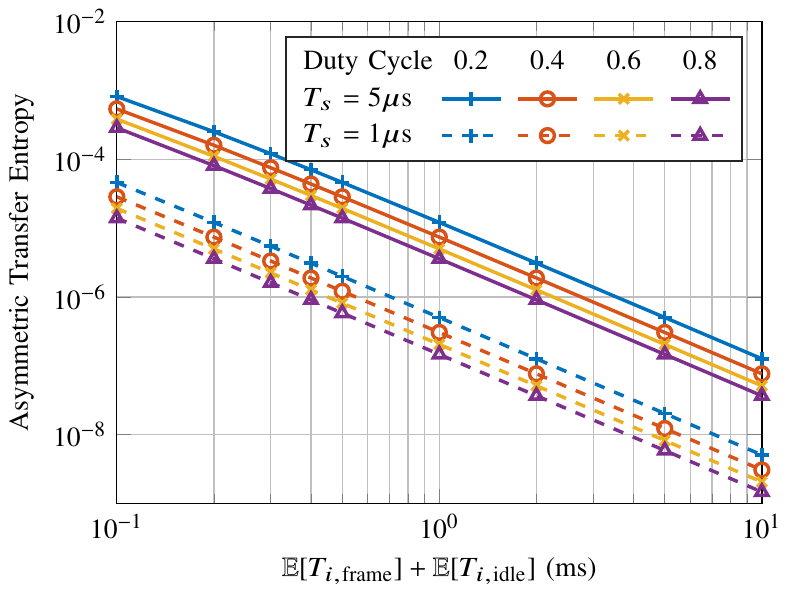}
    \caption{Theoretical asymmetric transfer entropy under the null hypothesis as the expected duration between the
    start of consecutive transmissions is varied}
    \label{fig:topology_th_pstart}
    \vspace*{-2em}
\end{figure}

One reason for studying this model is to understand how $A_{E_i \rightarrow S_j|E_j}(1)$ behaves under the null
hypothesis. For this purpose, we set $p_{r_j} = 0 = p_{r_i}$. As discussed above, the estimated response time for a null
hypothesis link will be 1. From Appendix~\ref{appendix:topology_ate_derivation}, we can express $A_{E_i \rightarrow
S_j|E_j}(1)$ as
\begin{align}
    A_{E_i \rightarrow S_j|E_j}(1) = & \frac{\rho_0 - 2p_j - p_i}{\rho_0} \log \left( 1 - \frac{p_i}{\rho_0 - 2p_j}
    \right) \notag \\
    & + \frac{\rho_0 - p_j}{\rho_0} \log\left( \frac{\rho_0 - p_j}{\rho_0 - p_j - p_i} \right)
    \label{eq:topology_atelag1}
\end{align}
where $\rho_0 = 1 + 4p_i + p_i(1 - p_{d_i})^{-1} + 4p_j + p_j(1 - p_{d_j})^{-1}$. Now, for the purposes of this study,
let us set $p_i = p_j$ and rewrite $p_{d_i}$ and $p_{d_j}$ using (\ref{eq:topology_pdi_def}) in the form of a duty
cycle percentage. Then, Fig.~\ref{fig:topology_th_pstart} shows that $A_{E_i \rightarrow S_j|E_j}(1)$ increases as the
frames become shorter and more frequent. It also shows that reducing the sampling interval $T_s$ reduces $A_{E_i
\rightarrow S_j|E_j}(1)$. Intuitively, this makes sense because a shorter $T_s$ implies more samples in the interval
between transmissions. We shall revisit (\ref{eq:topology_atelag1}) and Fig.~\ref{fig:topology_th_pstart} in
Section~\ref{sec:topology_simulations} for understanding the effect of increasing $N$.

\subsection{Effect of increasing response probability}
\label{sec:topology_th_probresp}

\begin{figure}
    \centering
    \footnotesize
    \includegraphics{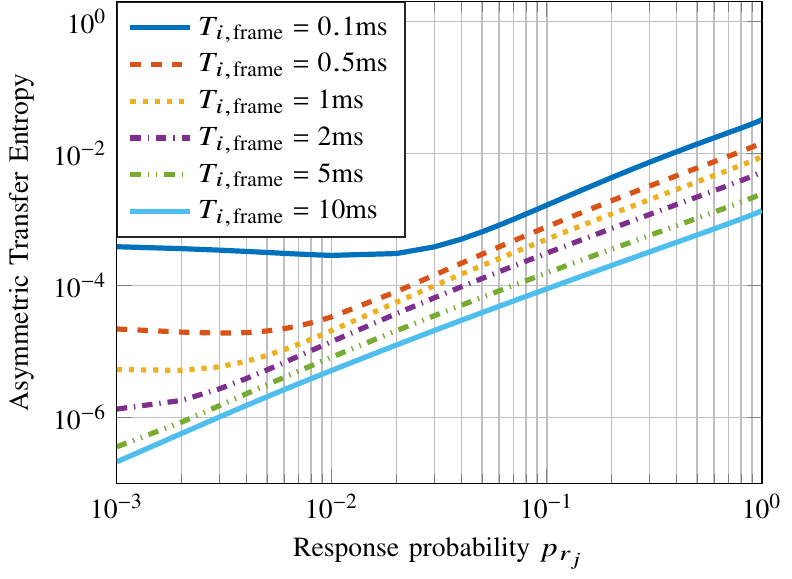}
    \caption{Asymmetric transfer entropy under alternate hypothesis for increasing response probability. Both IUs $i$
    and $j$ have equal probability of starting a new transmission and frame length distributions. Parameter: Response
    duration 250$\mu$s, mean idle time = 100$\mu$s.}
    \label{fig:topology_mc_link_response}
    \vspace*{-2em}
\end{figure}

As is to be expected, the asymmetric transfer entropy increases with the probability $p_{r_j}$ of IU $j$ responding to
IU $i$. Fig.~\ref{fig:topology_mc_link_response} shows that this is true only for a sufficiently large response
probability depending on the frame length when a constant idle time is maintained. Together with the discussion for
Fig.~\ref{fig:topology_th_pstart}, we infer that for short and frequent transmissions, the asymmetric transfer entropy
under the null hypothesis is equivalent to that for a small response probability. Hence,
Fig.~\ref{fig:topology_mc_link_response} further highlights the difficulty of detecting causality for IUs with short
frequent transmissions.

\section{Simulation Results}
\label{sec:topology_simulations}

In this section, we study the performance of learning the topology of 802.11n networks through NS-3
simulations\footnote{We used commit 578f6c0 from the ns-3-dev Git repository
(\texttt{https://github.com/nsnam/ns-3-dev-git}) in order to use the latest SpectrumWifiPhy implementation. The
MonitorSnifferTx trace was modified to obtain the physical layer transmit duration for each packet.}. Note that 802.11n
uses time division multiplexing opportunistic spectrum access through its Distributed Coordination Function (DCF). The
simulation uses a 20MHz wide channel in the 5GHz band with a datarate of 19.5Mbps. The response time (called SIFS) for
this setting is 16$\mu$s while the minimum interval (called DIFS) between the end of one transmission and the start of
another is 34$\mu$s. We use this set up for studying infrastructure-based networks of 2 access points (APs) located 40m 
apart and stations (STAs)
uniformly distributed in a disc of radius 15m. NS-3's OnOffApplications are set up for both uplink and downlink flows at
all STAs and APs.  These applications have exponentially distributed on and off times.  The parameters used for the NS-3
simulations of each of the following simulations are listed in Table~\ref{tab:topology_sim_params}.

For our Monte Carlo simulations, we present results for different infrastructure-based topologies after averaging over 
time.  That is, each topology is simulated for up to 1000 seconds and each of the topology learning algorithms are run 
on non-overlapping time windows of the obtained activity sequences. The length of these windows is described for each of 
the scenarios below. The learning algorithms were implemented in MATLAB for ease of development.  We measure the 
performance of the algorithms as the average fraction of links detected and the average number of extra links, i.e., 
pairs of IUs erroneously detected as being linked. We consider links without direction. 

In this section, we will use the Monte Carlo simulations to compare our proposed ATELNeT algorithm with the hard and
soft fusion algorithms proposed in~\cite{tilghman_inferring_2013}, the Hawkes process based method proposed
in~\cite{moore_analysis_2016}\footnote{The code for~\cite{moore_analysis_2016} was obtained from the authors' website:
\texttt{http://mdav.ece.gatech.edu/software/}}, and the linear asymmetric test proposed in
Section~\ref{sec:topology_proposed_linear}. Then, we will present a comparison to the algorithm proposed
in~\cite{kokalj-filipovic_learning_2016} for learning end-to-end routes in ad hoc networks. Finally, we will compare the
computational complexity of the different algorithms.

\subsection{Comparison of Parameters}
\label{sec:topology_sim_inputs}

\begin{table}
    \centering
    \caption{Simulation parameters for the simulations in Section~\ref{sec:topology_simulations}}
    \label{tab:topology_sim_params}
    \begin{tabular}{l r r r}
        \toprule
        Parameters             & ROC         & Duration  & Network Size \\
        \midrule
        Protocol               & 802.11n     & 802.11n   & 802.11n      \\
        Number of APs          & 2           & 2         & 2            \\
        Number of STAs per AP  & 3           & 3         & 1 to 6       \\
        Uplink mean on time    & 1ms         & 1ms       & 1ms          \\
        Uplink mean off time   & 10ms        & 10ms      & 10ms         \\
        Downlink mean on time  & None        & None      & 1ms          \\
        Downlink mean off time & None        & None      & 10ms         \\
        Observation duration   & 60ms, 600ms & 1ms to 5s & 1s           \\
        \bottomrule
    \end{tabular}
    \vspace*{-2.5em}
\end{table}

Though the input for each of the algorithms mentioned above are the same binary activity sequences~$a_m[n]$, each of
these methods uses a different sampling interval and causal lag. For the hard and soft fusion algorithms
of~\cite{tilghman_inferring_2013}, we input the activity sequences sampled at $T_s = 20\mu$s, used a window size of
60ms, and set the causal lag to 8 samples, i.e., 160$\mu$s as per the authors recommendations.
Since~\cite{moore_analysis_2016} studies the continuous time point process of the transmission start times, we input the
absolute start times of each transmission. A causal lag of 160$\mu$s was used via an exponentially decaying kernel.

For ATELNeT, we ensure $T_s \tau_{\max}$ is greater than the minimum idle time for any IU. In general, we want $T_s$ to
be shorter than the minimum response time for any pair of IUs and, as Fig.~\ref{fig:topology_th_pstart} indicates, $T_s$
should be as small as possible. For the 802.11 family, the minimum idle time varies from 34$\mu$s to 50$\mu$s. Hence,
we choose $T_s = 5\mu$s and $\tau_{\max} = 10$, i.e., maximum causal lag of 50$\mu$s. 

For the linear test for asymmetric Granger causality, we choose the same sampling interval $T_s = 5\mu$s, but set the
causal lag as 3 samples to match the response time of the 802.11n system being simulated.

\subsection{Comparison of Test Statistics}
\label{sec:topology_sim_roc}

\begin{figure}
    \centering
    \subfloat[Observed duration: 60ms]{
    \footnotesize
        \includegraphics{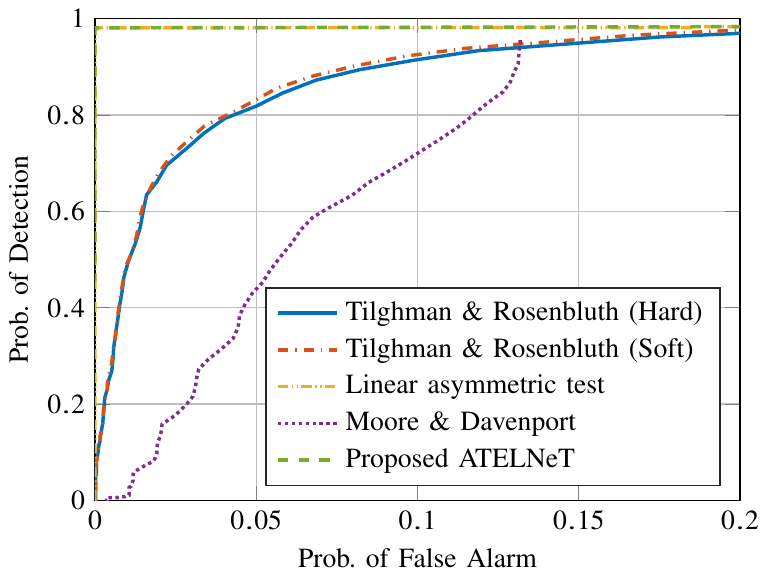}
        \label{fig:topology_roc_best_60ms}
    }
    \\
    \subfloat[Observed duration: 600ms]{
    \footnotesize
        \includegraphics{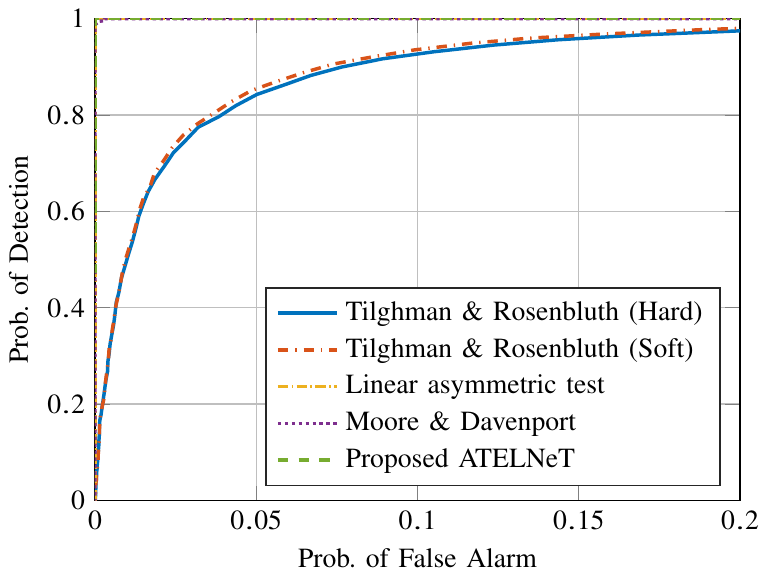}
        \label{fig:topology_roc_best_600ms}
    }
    \caption{ROC of the topology learning algorithm for a system of 2 APs 40m apart and 3 STAs each. For null
             hypothesis, test statistics for two STAs associated with different APs were used. For alternate hypothesis,
             test statistics for an AP and an associated STA were used. Test statistics were computed for 1000
             non-overlapping windows of duration 60ms in (a) and 600ms in (b).}
    \label{fig:topology_roc_best}
    \vspace*{-1.8em}
\end{figure}

We begin by studying the receiver operating characteristics (ROC) for the test statistics proposed
in~\cite{tilghman_inferring_2013,moore_analysis_2016} and Section~\ref{sec:topology_proposed}. For the hard fusion
algorithm proposed in~\cite{tilghman_inferring_2013}, we used $g_{i\rightarrow j}$ from
(\ref{eq:topology_tilghman_hard}) as the test statistic. For the soft fusion algorithm also proposed
in~\cite{tilghman_inferring_2013}, we used the causality magnitude $F_{i \rightarrow j}$ from
(\ref{eq:topology_causality_magnitude}). For the algorithm proposed in~\cite{moore_analysis_2016}, we used their
\emph{influence matrix} as the test statistic.

First, we used 60ms long observation periods to plot the ROC in Fig.~\subref*{fig:topology_roc_best_60ms}. We note that 
both test statistics proposed in~\cite{tilghman_inferring_2013} have almost identical curves. On the other hand, the 
algorithm proposed in~\cite{moore_analysis_2016} appears to have an upper limit on the achievable detection probability.  
Furthermore, since the authors do not propose a minimum strength required for detecting the causal connection, their 
threshold is chosen to be zero, i.e., they operate at the upper limit of the detection probability as seen in 
Fig.~\subref*{fig:topology_roc_best_60ms}. Hence, their proposed method suffers from a large false alarm rate as will be 
seen in the later results as well. Finally, test statistics of both ATELNeT, i.e., $\hat{A}_{E_i \rightarrow S_j|E_j}(N, 
\hat{\tau}_{i,j})$, and linear asymmetric test, i.e., (\ref{eq:topology_tilghman_hard}), have approximately the same 
ROC.

Next, Fig.~\subref*{fig:topology_roc_best_600ms} shows the ROC for observation periods of 600ms. Since both
algorithms proposed in~\cite{tilghman_inferring_2013} use windows of length 60ms, the ROC for their test
statistics do not change with increasing observation periods. However, the algorithm proposed
in~\cite{moore_analysis_2016} does improve its ROC making it a viable candidate for testing the binary hypotheses.
The linear asymmetric test as well as ATELNeT have almost identical and perfect ROC curves.

The perfect nature of the ROC curves in Fig.~\subref*{fig:topology_roc_best_600ms} indicates that any of the three methods
-- that proposed in~\cite{moore_analysis_2016} and those we propose in Section~\ref{sec:topology_proposed} -- should be
able to achieve almost perfect detection of the network topology by an appropriate choice of thresholds. However, the
simulation results below will show that the thresholds chosen for each algorithm are suboptimal. We believe this is due
to two facts: the threshold is chosen to have constant false alarm and the null hypothesis in all these works models
zero causality.

\subsection{Performance vs. Observation Duration}
\label{sec:topology_sim_durations}

\begin{figure}
\makebox[\subfloatwidth][c]{
    \subfloat[Average fraction of undirected links detected]{
    \footnotesize
        \includegraphics{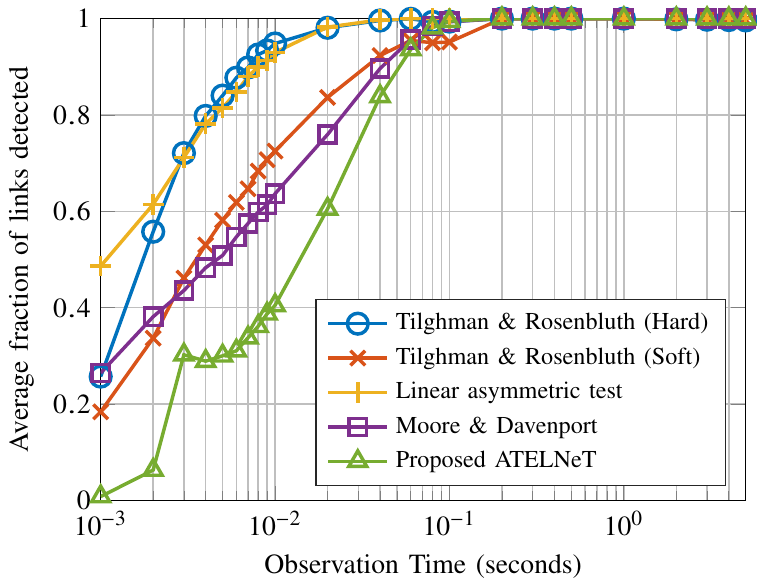}
        \label{fig:topology_sim_duration_det}
    }
}
\makebox[\subfloatwidth][c]{
    \subfloat[Average number of extra links detected]{
    \footnotesize
        \includegraphics{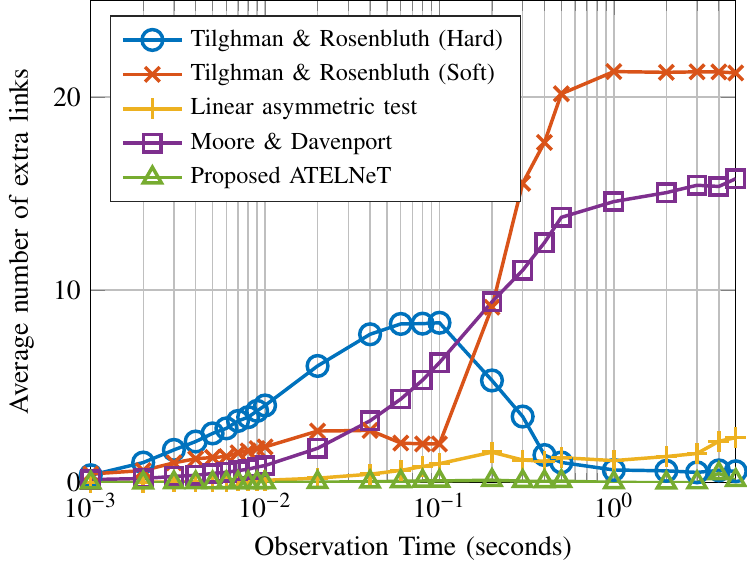}
        \label{fig:topology_sim_duration_extra}
    }
}
    \caption{Performance of inferring network topology as observation duration increases. System: 2 802.11n APs with 3
        STAs each.}
    \label{fig:topology_sim_duration}
    \vspace*{-1.7em}
\end{figure}

Increasing the observation period or, equivalently, the number of samples $N$ has multiple effects. First, the
non-stationary nature of the communication protocols means that more responses are observed on each link. Secondly, with
increasing number of samples, each of the methods is able to detect weaker causal relationships. Hence, the false alarm
rate of each of the methods, including ATELNeT, increases with increasing observation period or number of samples.

Fig.~\ref{fig:topology_sim_duration} shows the performance of each of these algorithm in learning the topology of the
system with 2 APs and 3 STAs associated with each AP. As expected, the fraction of links detected correctly increases
with the observation period. The higher detection rate of the linear tests is also associated with a higher number of
extra links as seen in Fig.~\subref*{fig:topology_sim_duration_extra}.  Both algorithms proposed
in~\cite{tilghman_inferring_2013} attempt to reduce the number of extra links by splitting the observed period into
windows of equal duration and fusing the topologies inferred on each window. As seen in
Fig.~\subref*{fig:topology_sim_duration_extra}, the hard fusion algorithm of~\cite{tilghman_inferring_2013} is more
effective at reducing the number of extra links detected than the soft fusion algorithm. Note that the maximum number of
extra links detected by the hard fusion algorithm is at about 60ms, i.e., the duration of a single window.

The proposed ATELNeT algorithm detects the least number of extra links. Since the linear asymmetric test also detects
significantly fewer extra links than the methods of~\cite{tilghman_inferring_2013}, we believe this performance gain is
due to testing asymmetric Granger causality instead of symmetric Granger causality. Note that the linear asymmetric test
does not use windowing or fusion.

We can also see this rise from our model using (\ref{eq:topology_atelag1}). By the non-central $\chi^2$ distribution of
$\hat{A}_{E_i \rightarrow S_j|E_j}(N, 1)$, we can write the false alarm probability $P_{\text{FA}}$ as
\begin{equation}
    P_{\text{FA}} = Q_1\left(\sqrt{2(N - 1)A_{E_i \rightarrow S_j|E_j}(1)}, \sqrt{\lambda(1)}\right) \label{eq:mc_pfa}
\end{equation}
where $Q_1(\cdot, \cdot)$ is Marcum's Q function. Hence, $P_{\text{FA}}$ increases with $N$ due to the monotonic nature
of the Marcum's Q function. Now, our Markov chain model has more dependencies than the actual
system due to assumptions such as lack of collisions. Therefore, (\ref{eq:mc_pfa}) provides an upper bound for
$P_{\text{FA}}$ rather than an equality. For example, the system considered in Fig.~\ref{fig:topology_sim_duration} has
$A_{E_i \rightarrow S_j|E_j}(1)$ of the order of $10^{-7}$ under the null hypothesis as per
Fig.~\ref{fig:topology_th_pstart}. For a 5s observation period, i.e., $N = 10^6$, (\ref{eq:mc_pfa}) tells us that the
false alarm probability is bounded above by 0.0653 while the simulation resulted in a false alarm probability of 0.0083.

\subsection{Performance as Network Size Increases}
\label{sec:topology_sim_nsize}

\begin{figure}
\makebox[\subfloatwidth][c]{
    \subfloat[Average fraction of undirected links detected]{
    \footnotesize
    \includegraphics{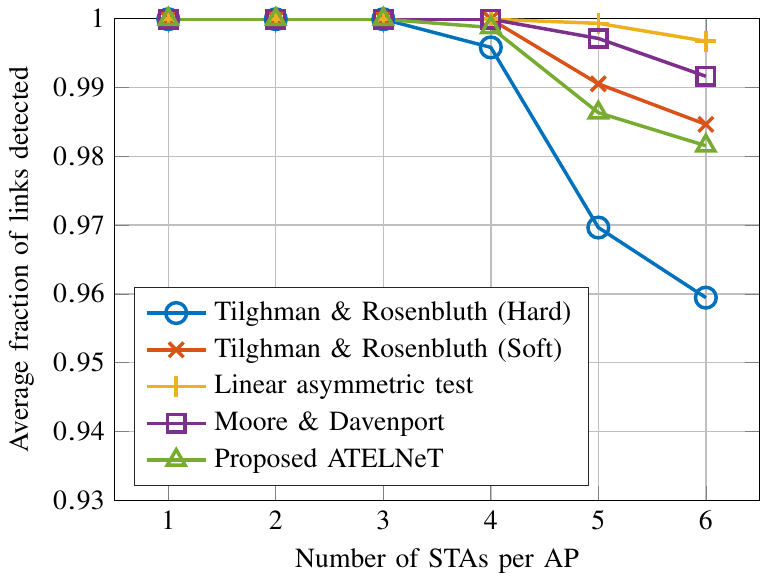}
    \label{fig:topology_sim_numSTAs_det}
    }
}
\makebox[\subfloatwidth][c]{
    \subfloat[Average number of extra links detected]{
    \footnotesize
    \includegraphics{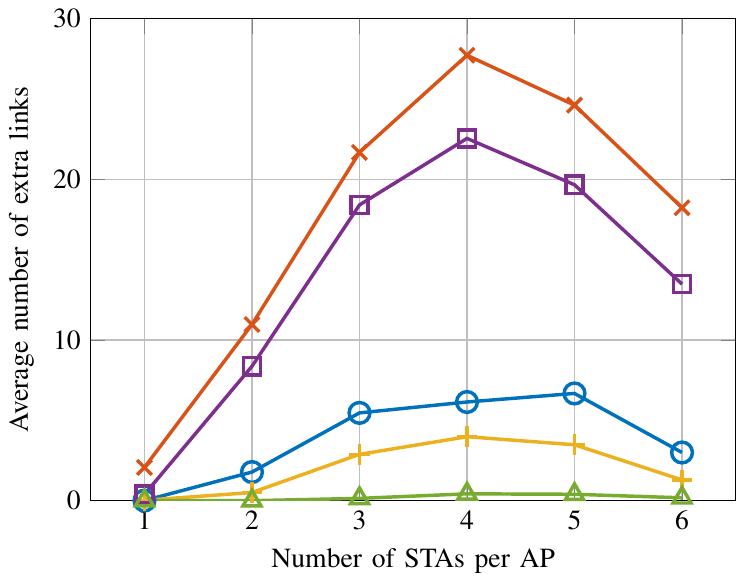}
    \label{fig:topology_sim_numSTAs_extra}
    }
}
    \caption{Performance of inferring network topology as the number of STAs increases. System: 2 802.11n APs with up to 6
        STAs each.}
    \label{fig:topology_sim_numSTAs}
    \vspace*{-2em}
\end{figure}

\newlength\unit
\setlength\unit{0.15cm}
\begin{figure*}
    \centering
        \subfloat[At 35s]{
            \scriptsize
            \includegraphics{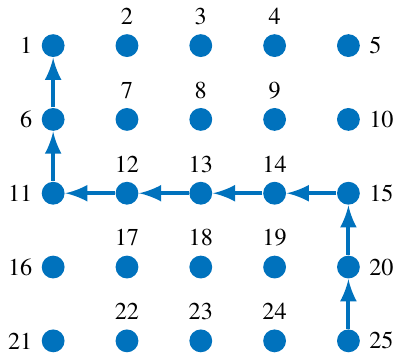}
            \label{fig:topology_adhoc1}
        }
        ~~~~~
        \subfloat[At 40s]{
            \scriptsize
            \includegraphics{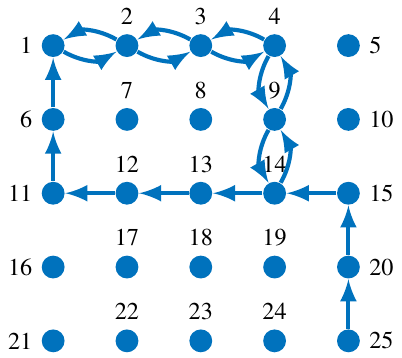}
            \label{fig:topology_adhoc2}
        }
        ~~~~~
        \subfloat[At 41s]{
            \scriptsize
            \includegraphics{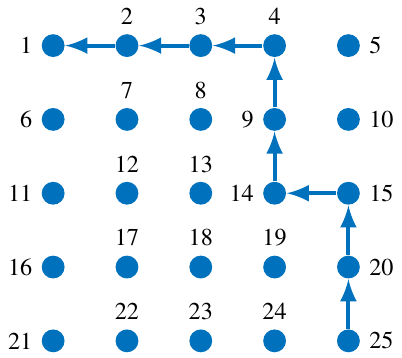}
            \label{fig:topology_adhoc3}
        }
        \\
        \subfloat[At 43s]{
            \scriptsize
            \includegraphics{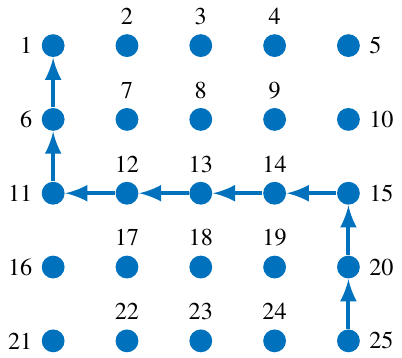}
            \label{fig:topology_adhoc1b}
        }
        ~~~~~
        \subfloat[At 111s]{
            \scriptsize
            \includegraphics{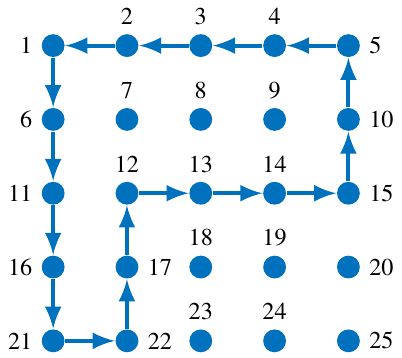}
            \label{fig:topology_adhoc4}
        }
        ~~~~~
        \subfloat[Output of~\cite{kokalj-filipovic_learning_2016} over full 150s]{
            \scriptsize
            \adjincludegraphics[width=0.26\textwidth,trim={0 0 {.405\width} 0},clip]{adhoc_output.tif}
            \label{fig:topology_adhoc_op}
        }
    \caption{Links learned from activity sequences of 25 IU ad hoc network. Data flow was set up from IU 25 to IU 1 from
    time 30s to 80s and from IU 21 to IU 5 from time 110s to 140s. 1s long observation periods were used for learning
    network topology and activity sequences were sampled at $T_s = 5\mu$s. Each link has 1Mbps data rate and 9$\mu$s
    response time.}
    \label{fig:topology_adhoc}
    \vspace*{-2em}
\end{figure*}

In Fig.~\ref{fig:topology_sim_numSTAs}, we compare the performance of learning the network topology as the number of
STAs associated with each AP increase. The trends are similar to that seen in~Fig.~\ref{fig:topology_sim_duration}. The
linear asymmetric test detects the highest fraction of links with a slightly higher number of extra links as compared to
ATELNeT. The algorithm of~\cite{moore_analysis_2016} and the soft fusion algorithm of~\cite{tilghman_inferring_2013}
detect a high number of extra links. The hard fusion algorithm of~\cite{tilghman_inferring_2013} detects a large
fraction of links correctly while detecting relatively fewer extra links. In summary, the ATELNeT
learns the topology of varying network sizes with high probability and with almost no extra links.

\subsection{Application: Inferring links in ad hoc networks}
\label{sec:topology_app_adhoc}

In ad hoc networks, the topology also corresponds to the routes of the data flow. Hence, the authors 
of~\cite{kokalj-filipovic_learning_2016} learn the end-to-end routes in ad hoc networks. For each IU, they propose a 
hidden semi-Markov model such that it has super states reflecting data flows that it participates in.  Their proposed 
algorithm learns the super state dependent distributions of the frame lengths and inter-arrival times for each IU using 
a hierarchical Dirichlet process as a prior for their super state model. IUs that have learned similar distributions are 
clustered together as part of the same end-to-end route.

In this section, we describe the directed links detected by our algorithm from the same data used
in~\cite{kokalj-filipovic_learning_2016}\footnote{Permission to use their data provided by the lead author
of~\cite{kokalj-filipovic_learning_2016}}. They simulated an ad hoc 802.11 network of 25 IUs located on a rectangular
grid such that only neighbouring IUs can communicate with each other. Constant bit rate UDP packets are used for data
flows routed using OLSR. From time 30s to 80s, 5 packets per second are sent from IU 25 to IU 1.  From time 110s to
140s, 5 packets per second are sent from IU 21 to IU 5.

From the results of ATELNeT shown in Fig.~\ref{fig:topology_adhoc}, we can infer that there were two routes for the data
flow from IU 25 to IU 1. At time 41s, we detect the route renegotiation and note that the new route remains in service
for at most 2 seconds since the original route is active from 43s onwards.  Fig.~\subref*{fig:topology_adhoc4} shows the
links detected for the data flow from IU 21 to IU 5. Interestingly, ATELNeT also detects links corresponding to a data
flow from IU 5 to IU 21 along a separate route that~\cite{kokalj-filipovic_learning_2016} had missed.

In comparison, Fig.~\subref*{fig:topology_adhoc_op} shows that the end-to-end routing algorithm proposed 
in~\cite{kokalj-filipovic_learning_2016} learns 3 routes (shown by distinct colors): one between IU 25 and IU 1 and two 
between IU 21 and IU 5. It does not detect the short lived route of Fig.~\subref*{fig:topology_adhoc2} nor does it 
detect the direction of the routes. Hence, our algorithm infers the directed links between pairs of IUs at a finer time 
resolution than~\cite{kokalj-filipovic_learning_2016} and can detect network changes faster.

\subsection{Comparison of Computational Complexity}

\begin{table}
    \centering
    \caption{Comparison of computational complexity of topology learning algorithms}
    \begin{tabular}{l p{10em} r}
        \toprule
        Algorithm                                       & Dominating Operation         & Complexity        \\
        \midrule
        ATELNeT                                         & Estimating joint prob. or testing large lags & 
        $O((N + 4^{\tau_{\max}})M^2)$ \\
        Hard fusion from~\cite{tilghman_inferring_2013} & Linear MMSE estimator        & $O(\tau^2 N M^2)$ \\
        Soft fusion from~\cite{tilghman_inferring_2013} & Linear MMSE estimator        & $O(\tau^2 N M^2)$ \\
        Linear asymmetric test                          & Linear MMSE estimator        & $O(\tau^2 N M^2)$ \\
        Moore, Davenport~\cite{moore_analysis_2016}   & Convex optimization          & $O(\tau_{\max} N M^2)$      
        \\
        \bottomrule
    \end{tabular}
    \vspace*{-2em}
\end{table}

For a single link, ATELNeT begins by estimating the joint probability mass function of ($S_j[t], E_i^{(\tau_{\max})}[t],
E_j^{(\tau_{\max})}[t]$) which requires $O(N)$ computations for $N$ samples. The joint probabilities for smaller lags
can be computed with $O(2^{\tau_{\max}})$ operations. Next, computing $A_{E_i \rightarrow S_j|E_j}(\tau)$ requires
$O(4^{\tau})$ operations. Hence, computing the asymmetric transfer entropy for $\tau_{\max}$ lags requires
$O(4^{\tau_{\max}})$ operations. In summary, the computational complexity of ATELNeT for a single link is
$O(N + 4^{\tau_{\max}})$. Testing all IU pairs requires $O((N + 4^{\tau_{\max}}) M^2)$ computations. The exponential 
rise in complexity due to $\tau_{\max}$ further reinforces the need for a better estimation of the maximum lag (in 
seconds) and the sampling period $T_s$.

In comparison, computing the linear MMSE estimator from $W$ samples for the parameters in a linear regression with
$\tau$ causal lag requires $O(\tau^2 W)$ computations. Hence, both fusion methods proposed
in~\cite{tilghman_inferring_2013} have computational complexity $O(\tau^2 W (N / W)) = O(\tau^2 N)$. Similarly, the
proposed linear asymmetric test has the same computational complexity. For testing all pairs of IUs, the computational
complexity of each algorithm is $O(\tau^2 N M^2)$. Note that these algorithms need an a priori accurate estimate of
$\tau$ which will increase the total computational complexity.

The method proposed in~\cite{moore_analysis_2016} solves $M$ convex optimization problems with a quasi-Newton descent
algorithm. Each of these problems has $M+1$ parameters and requires $O(\tau_{\max} NM)$ operations for evaluating 
their likelihood function. Therefore, each problem has $O((M+1)^2 + \tau_{\max} NM)$ computations.  For solving 
all $M$ problems, the computational complexity is $O(\tau_{\max} N M^2)$ since $M \ll N$ typically.

In summary, the computational complexity of ATELNeT can be greater than those of algorithms proposed
in~\cite{tilghman_inferring_2013} and~\cite{moore_analysis_2016} for large $\tau_{\max}$.

\section{Conclusion and Future Work}
\label{sec:topology_conclusion}

In this paper, we have proposed a method to learn the network topology of time multiplexing communicating IUs by
detecting the causal relationship of two IUs' transmissions due to their response time, i.e., the time between the end
of a transmitted frame and the start of the response signal. This response time is widely used in communication
protocols today and is relatively invariant over time for a given link. For detecting this response, we have proposed a
non-parametric test statistic called asymmetric transfer entropy that differs from existing methods in two ways: it
tests asymmetric Granger causality and does not require a linear model. We have shown that the shared channel causes IUs
to have non-zero causality irrespective of whether they are communicating with each other. Since this results in
detecting extra links, we minimized this effect by an algorithm that estimates the response time of the link.

We also proposed a Markov chain model to analyze the behaviour of the proposed test statistic under both hypotheses for
the transmissions of a pair of IUs. The model shows that both the false alarm and detection probability increase with
shorter more frequent transmissions. Using NS-3 simulations of 802.11n networks, we showed that, in comparison to
existing methods in literature, our proposed algorithm significantly reduces the number of extra links detected in the
network while achieving comparable link detection performance. Finally, we showed that the topology learned from an ad
hoc network actually consists of the directed links in the network.

This is an interesting problem that may be amenable to many approaches. Forward-backward iterative algorithms, graphical 
approaches to inferring the entire network simultaneously, as well as supervised learning based approaches may be of 
interest for future study. It would also be beneficial to study the incorporation of the learned topology into channel 
access protocols for cognitive radios.

\section{Acknowledgement}

The authors would like to thank Silvija Kokalj-Filipovic for sharing data from~\cite{kokalj-filipovic_learning_2016} and
interesting discussions.

\appendices
\section{Asymptotic Distribution of Empirically Estimated Asymmetric Transfer Entropy}
\label{appendix:topology_ate_distribution}

We now derive the distribution of the empirical estimate $\hat{A}_{E_i \rightarrow S_j|E_j}(N, \tau)$ using the same
logic as the derivation of the distribution of the empirical estimate transfer entropy in~\cite{barnett_transfer_2012}.
In short, $A_{E_i \rightarrow S_j|E_j}(\tau)$ is reinterpreted as a log likelihood ratio and then large sample theory is
used to show that the empirical estimator has a known distribution.

Since we do not have a specific model for the time series
$S_j[t]$, $E_i[t]$, and $E_j[t]$, the non-parametric estimator of $A_{E_i \rightarrow S_j|E_j}$ is computed from the
empirically estimated probability mass functions $\hat{P}(S_j[t], E_i^{(\tau)}[t], E_j^{(\tau)}[t])$, $\hat{P}(S_j[t]|
E_i^{(\tau)}[t], E_j^{(\tau)}[t])$, $\hat{P}(S_j[t], E_i^{(\tau)})$, and $\hat{P}(S_j[t]|E_j^{(\tau)}[t])$:
\begin{align}
&\hat{A}_{E_i \rightarrow S_j|E_j}(N, \tau) \notag \\
&=  -\sum_{s_j, e_j^{(\tau)}} \left[ \hat{P}\left(S_j[t] = s_j, E_j^{(\tau)}[t] =
e_j^{(\tau)}\right) \right. \notag \\
&\hspace{4em} \left. \times \log \hat{P}\left(S_j[t] = s_j | E_j^{(\tau)}[t] = e_j^{(\tau)}\right) \right] \notag \\
& + \sum_{s_j, e_j^{(\tau)}, e_i^{(\tau)}} \left[ \hat{P}\left(S_j[t] = s_j, E_j^{(\tau)}[t] = e_j^{(\tau)}, E_i^{(\tau)}[t] =
e_i^{(\tau)}\right) \right. \notag \\
&\hspace{4em} \left. \times \log \hat{P}\left(S_j[t] = s_j | E_j^{(\tau)}[t] = e_j^{(\tau)}, E_i^{(\tau)}[t] =
e_i^{(\tau)}\right) \right] \notag
\end{align}
where $s_j \in \{0,1\}$, $e_j^{(\tau)} \in \{0,1\}^{\tau}$, and $e_i^{(\tau)} \in \{0,1\}^{\tau}$. These empirical
probability mass functions are also maximum likelihood estimators of the actual probability mass functions and can be
considered as the parameters for the model for $S_j[t]$ given $E_i^{(\tau)}[t]$ and $E_j^{(\tau)}[t]$. Then, we can
write the likelihood function for the model as
\begin{align}
    \mathcal{L}\left(\mathbf{\theta}|S_j[t], E_j^{(\tau)}[t], E_i^{(\tau)}[t]\right) = & P\left(S_j[t]|E_j^{(\tau)}[t],
    E_i^{(\tau)}[t]; \mathbf{\theta}\right) \notag \\
    & \times P\left(E_j^{(\tau)}[t], E_i^{(\tau)}[t]\right)
    \label{eq:topology_ate_likelihood}
\end{align}
where $\mathbf{\theta}$ is the model parameter vector. Now, under our null hypothesis, $S_j[t]$ is conditionally
independent of $E_i^{(\tau)}[t]$ given $E_j^{(\tau)}[t]$. Let $\Theta_0$ be the set of parameter vectors under this null
hypothesis and $\Theta_1$ be the set of parameter vectors under the alternate hypothesis. The likelihood ratio is given by
\begin{align}
&\Lambda\left(S_j^{(N)}[N + 1], E_i^{(N)}[N + 1], E_j^{(N)}[N+1] \right) \notag \\
&\qquad \equiv \frac{\mathcal{L}\left(\hat{\mathbf{\theta}}_0 | S_j^{(N)}[N + 1], E_i^{(N)}[N + 1],
E_j^{(N)}[N+1]\right)}{\mathcal{L}\left(\hat{\mathbf{\theta}}_1 | S_j^{(N)}[N + 1], E_i^{(N)}[N + 1], E_j^{(N)}[N+1]\right)}
    \label{eq:topology_ate_lr}
\end{align}
where $\hat{\mathbf{\theta}}_0$ and $\hat{\mathbf{\theta}}_1$ are the maximum likelihood estimators of $\mathbf{\theta}$
under the two hypotheses. Note that as mentioned above, these are simply the empirical probability mass functions of
$S_j[t], E_j^{(\tau)}[t]$ and $S_j[t], E_j^{(\tau)}[t], E_i^{(\tau)}[t]$ respectively. Now, it is easy to see that
(\ref{eq:topology_ate_lr}) combined with (\ref{eq:topology_ate_likelihood}) gives
\begin{align}
& (N - \tau) \hat{A}_{E_i \rightarrow S_j|E_j}(N, \tau) \notag \\
& \qquad \equiv \log \Lambda \left(S_j^{(N)}[N+1], E_j^{(N)}[N+1], E_i^{(N)}[N+1]\right)
\end{align}

Hence, large sample theory~\cite[Theorem IX, Pg. 480]{wald_tests_1943} tells us that under the null hypothesis, the
distribution of $2(N - \tau) \hat{A}_{E_i \rightarrow S_j | E_j}(N, \tau)$ is asymptotically a central $\chi^2$ 
distribution with
${d_{i,j}}$ degrees of freedom while under the alternate hypothesis, the distribution of $2(N - \tau) \hat{A}_{E_i
\rightarrow S_j | E_j}(N, \tau)$ is asymptotically a non-central $\chi^2$ distribution with $d_{i,j}$ degrees of freedom 
and
non-centrality parameter $2(N-\tau) A_{E_i \rightarrow S_j|E_j}(\tau)$. Here, $d_{i,j}$ is the difference between the
number of parameters in the full and null models.

\section{Steady state probabilities of Markov Chain}
\label{appendix:topology_steady_state}

\begin{figure*}[!t]
\setcounter{mytempeqncnt}{\value{equation}}
\setcounter{equation}{41}
\begin{align}
&A_{E_i \rightarrow S_j|E_j}(1) = (1 - \pstar_{js} - \pstar_{ch_j[1]} - \pstar_{ch_{ij}[1]} - \pstar_{jr_s} -
\pstar_{ch_i[1]} - \pstar_{ch_{ji}[1]}) \notag \\
& \hspace{7em} \times \log \left[ \frac{(1 - \pstar_{js} - \pstar_{ch_j[1]} - \pstar_{ch_{ij}[1]} - \pstar_{jr_s} -
    \pstar_{ch_i[1]} - \pstar_{ch_{ji}[1]}) P(E_j[t-1] = 0)}{P(E_i[t-1] = 0, E_j[t-1]= 0)(1 - \pstar_{js} -
\pstar_{ch_j[1]} - \pstar_{ch_{ij}[1]} - \pstar_{jr_s})} \right] \notag \\
&\hspace{1.5em} + (\pstar_{js} + \pstar_{jr_s}) \log \left[ \frac{(\pstar_{js} + \pstar_{jr_s}) P(E_j[t-1] = 0)}{P(E_j[t-1] = 0,
E_i[t-1] = 0) (\pstar_{js} + \pstar_{jr_s})} \right]  + (\pstar_{ch_i[1]} + \pstar_{ch_{ji}[1]}) \log \frac{P(E_j[t-1]=0)}{P(S_j[t]=0,E_j[t-1]=0)}
\label{eq:topology_ate_lag1_raw}
\end{align}
\setcounter{equation}{\value{mytempeqncnt}}
\hrulefill
\end{figure*}

In this section, we derive the steady state probabilities of the Markov chain model described in
Section~\ref{sec:topology_analysis} for a two IU system. First, we note that all the steady state probabilities in this
model can be written in terms of $\pstar_{is}$ and $\pstar_{js}$:
\begin{align}
\pstar_i            & = \pstar_{is}(1 - p_{d_i})^{-1} \label{eq:topology_ss_subfirst}\\
\pstar_{ch_i[2]}    & = \pstar_{ch_i[1]} = \pstar_{is} \\
\pstar_{ch_i[3]}    & = (1 - p_{r_j}) \pstar_{is} \\
\pstar_{jr_s}       & = p_{r_j} \pstar_{is} \\
\pstar_{jr}         & = p_{r_j}(1 - p_{dr_j}) \pstar_{is} \\
\pstar_{ch_{ij}[3]} & = \pstar_{ch_{ij}[2]} = \pstar_{ch_{ij}[1]} = p_{r_j} \pstar_{is} \\
\pstar_j            & = \pstar_{js}(1 - p_{d_j})^{-1} \\
\pstar_{ch_j[2]}    & = \pstar_{ch_j[1]} = \pstar_{js} \\
\pstar_{ch_j[3]}    & = (1 - p_{r_i}) \pstar_{js} \\
\pstar_{ir_s}       & = p_{r_i} \pstar_{js} \\
\pstar_{ir}         & = p_{r_i}(1 - p_{dr_i}) \pstar_{js} \\
\pstar_{ch_{ji}[3]} & = \pstar_{ch_{ji}[2]} = \pstar_{ch_{ji}[1]} = p_{r_i} \pstar_{js}. \label{eq:topology_ss_sublast}
\\
\pstar_{ch[\infty]} &= (1 - p_i - p_j) \pstar_{ch[\infty]} + \pstar_{ch_i[3]} + (1 - p_i -
p_j)\pstar_{ch_{ij}[3]} \notag \\
&\quad + (1 - p_i - p_j)\pstar_{ch_{ji}[3]} + \pstar_{ch_j[3]} \label{eq:topology_pstar_chinfty} \\
&= \left( \frac{1}{p_i + p_j} - p_{r_j} \right) \pstar_{is} + \left( \frac{1}{p_i + p_j} -
p_{r_i} \right) \pstar_{js} \label{eq:topology_ss_sub1}
\end{align}
Hence, the only unknown variables are $\pstar_{is}$ and $\pstar_{js}$. Using (\ref{eq:topology_ss_sub1}) and the fact
that
\begin{align}
\pstar_{is}         & = p_i \pstar_{ch[\infty]} + p_i \pstar_{ch_{ij}[3]} + p_i \pstar_{ch_{ji}[3]}
\label{eq:topology_pstar_is} \\
&= p_i \pstar_{ch[\infty]} + p_i p_{r_j} \pstar_{is} + p_i p_{r_i} \pstar_{js}
\label{eq:topology_pstar_is_simplified}
\end{align}
we get that
\begin{equation}
    \frac{p_i}{\pstar_{is}} = \frac{p_j}{\pstar_{js}} \triangleq \rho.
    \label{eq:topology_pstar_ratios}
\end{equation}
Then, enforcing the sum of the steady state probabilities of all the states to be 1 gives us
\begin{align}
    \rho = 1 & + 4p_i + 2p_i p_{r_j} + \frac{p_i}{1 - p_{d_i}} + \frac{p_{r_j}p_i}{1 - p_{d_{r_j}}}
    \notag \\
    & + 4p_j + 2p_j p_{r_i} + \frac{p_j}{1 - p_{d_j}} + \frac{p_{r_i}p_j}{1 - p_{d_{r_i}}}.
    \label{eq:topology_rho_def}
\end{align}

\section{Derivation of asymmetric transfer entropy from model for various causal lags}
\label{appendix:topology_ate_derivation}

\subsection{Causal lag 1}

If the lag ($\tau$) is 1, then we get
\begin{align}
&P(E_j[t - 1] = 0) = 1 - p_j\rho^{-1} - p_{rj}p_i \rho^{-1} \quad \text{and} \\ %
& P(E_j[t - 1] = 0, E_i[t - 1] = 0) \notag \\
& \qquad = 1 - (1 + p_{ri}) p_j\rho^{-1} - (1 + p_{rj}) p_i \rho^{-1}.
\end{align}
Now we can write $A_{E_i \rightarrow S_j|E_j}(1)$ as shown in (\ref{eq:topology_ate_lag1_raw}).
\addtocounter{equation}{1}
Simplifying (\ref{eq:topology_ate_lag1_raw}) using (\ref{eq:topology_ss_subfirst})-(\ref{eq:topology_ss_sub1}) and
(\ref{eq:topology_rho_def}), we get (\ref{eq:topology_ate_lag1}).
\begin{align}
&\rho A_{E_i \rightarrow S_j|E_j}(1) = \left[ \rho - p_j(2 + p_{r_i}) - p_i (1 + 2p_{r_j}) \right] \notag \\
& \hspace{7em} \times \log \left[ 1 - \frac{p_i + p_{r_i}p_j}{\rho - 2p_j - 2p_{r_j}p_i} \right] \notag \\
&+ (\rho - p_j - p_{r_j}p_i) \log \left[ \frac{\rho - p_j - p_{r_j}p_i}{\rho - p_j(1+p_{r_i}) - p_i(1 +
p_{r_j})}\right]
\label{eq:topology_ate_lag1}
\end{align}

\subsection{Causal lag 2}

Similarly, we can write the asymmetric transfer entropy with causal lag 2 as follows.
\begin{align}
&\rho A_{E_i \rightarrow S_j|E_j}(2) \notag \\
&= \left\{ \rho - p_j(3 +2p_{r_i}) - p_i(2+3p_{r_j}) \right\} \notag \\
& \times \log \left\{ \frac{ \rho - p_j(3 +2p_{r_i}) - p_i(2+3p_{r_j}) }{( \rho - 2p_j -
2p_{r_j}p_i - 2p_i - 2p_{r_i}p_j )  } \frac{ \rho - 2p_j - 2p_{r_j}p_i }{ \rho - 3p_j - 3p_{r_j}p_i } \right\} \notag \\
& + (p_j + p_{r_j}p_i) \log \frac{\rho - 2p_j - 2p_{r_j}p_i}{\rho - 2p_j - 2p_{r_j}p_i - 2p_i - 2p_{r_i}p_j} \notag \\
& + \left[ p_i(2 + p_{r_j}) + p_jp_{r_i} \right] \log\frac{ \rho - 2p_j - 2p_{r_j}p_i }{ \rho - 3p_j - 3p_{r_j}p_i } 
\end{align}

\subsection{Causal lag 3}

Before expressing $A_{E_i \rightarrow S_j|E_j}(3)$, we note a few preliminaries for convenience.
\begin{align}
& P(E_i^{(3)}[t]=(0,0,0), E_j^{(3)}[t] = (0,0,0)) \notag \\
&= \rho^{-1}\left\{ 1 - p_i p_{r_j} - p_j p_{r_i} + p_i(1 - p_{d_i})^{-1} + p_i p_{r_j}(1 - p_{dr_j}) \right. \notag \\
& \hspace{3.5em} \left. + p_j p_{r_i}(1 - p_{dr_i}) + p_j(1 - p_{d_j})^{-1}  + p_j + p_i \right\}, \\
&P(E_j^{(3)}[t] = (0, 0, 0)) 
= 1 - p_j \rho^{-1} (3 + 2p_{r_i}) \text{, and} \\
&P(S_j[t] = 0, E_j^{(3)}[t] = (0,0,0)) \notag \\
&= \rho^{-1} \left\{ 1 + p_i\left[ 4 + (1 - p_{d_i})^{-1} - p_{r_j}(1 + p_{dr_j})\right] \right. \notag \\
& \hspace{4em} \left. + p_j \left[ (1 - p_{d_j})^{-1} + p_{r_i} (3 - p_{dr_i}) \right] \right\}.
\end{align}
Then, we can write $A_{E_i \rightarrow S_j|E_j}(3)$ after simplification as:
\begin{align}
&\rho A_{E_i \rightarrow S_j|E_j}(3) \notag \\
&= \left[\rho -p_j(3 + 2p_{r_i})\right] \log \left[1 - p_j\rho^{-1}(3 +
2p_{r_i})\right] \notag \\
& + \rho_3 \log \left\{ \frac{\rho_3}{\rho P(E_i^{(3)}[t] = (0,0,0), E_j^{(3)}[t] = (0,0,0))} \right\} \notag \\
& - \rho_3 \log \left\{ P(S_j[t] = 0, E_j^{(3)}[t] = (0,0,0)) \right\} \notag \\
& + p_j \log \frac{p_j}{(p_j + p_{r_j}p_i) P(E_i^{(3)}[t] = (0,0,0), E_j^{(3)}[t] = (0,0,0))} \notag \\
& + \left[ (1 - p_{r_j})p_i + p_{r_i} p_j \right] \notag \\
& \qquad \times \log \frac{p_i(1 - p_{r_j}) + p_{r_i}p_j}{(p_i + p_{r_i}p_j) P(S_j[t]
= 0, E_j^{(3)}[t] = (0,0,0))} \notag \\
& + p_{r_j} p_i \log \frac{\rho p_{r_j}p_i}{(p_i + p_{r_i}p_j)(p_j + p_{r_j}p_i)} \notag \\
& - 2(p_i + p_{r_i}p_j) \log P(S_j[t] = 0, E_j^{(3)}[t] = (0,0,0)).
\end{align}
where $\rho_3 \triangleq 1 + p_i\left[ 1 + (1 - p_{d_i})^{-1} + p_{r_j}\left((1 - p_{dr_j})^{-1} - 1\right) \right] + p_j \left[ (1 -
p_{d_j})^{-1} + p_{r_i} \left( (1 - p_{dr_i})^{-1} - 1 \right) \right]$.

\bibliographystyle{IEEEtran}
\bibliography{IEEEabrv,Learning_Network_Topology}

\begin{thebibliography}{10}
\providecommand{\url}[1]{#1}
\csname url@samestyle\endcsname
\providecommand{\newblock}{\relax}
\providecommand{\bibinfo}[2]{#2}
\providecommand{\BIBentrySTDinterwordspacing}{\spaceskip=0pt\relax}
\providecommand{\BIBentryALTinterwordstretchfactor}{4}
\providecommand{\BIBentryALTinterwordspacing}{\spaceskip=\fontdimen2\font plus
\BIBentryALTinterwordstretchfactor\fontdimen3\font minus
  \fontdimen4\font\relax}
\providecommand{\BIBforeignlanguage}[2]{{%
\expandafter\ifx\csname l@#1\endcsname\relax
\typeout{** WARNING: IEEEtran.bst: No hyphenation pattern has been}%
\typeout{** loaded for the language `#1'. Using the pattern for}%
\typeout{** the default language instead.}%
\else
\language=\csname l@#1\endcsname
\fi
#2}}
\providecommand{\BIBdecl}{\relax}
\BIBdecl

\bibitem{ali_advances_2017}
A.~Ali and W.~Hamouda, ``Advances on {{Spectrum Sensing}} for {{Cognitive Radio
  Networks}}: {{Theory}} and {{Applications}},'' \emph{IEEE Commun. Surv.
  Tutor.}, vol.~19, no.~2, pp. 1277--1304, Secondquarter 2017.

\bibitem{ieee_80222_2011}
``{{IEEE Standard}} for {{Information}} technology\textendash{} {{Local}} and
  metropolitan area networks\textendash{} {{Specific}}
  requirements\textendash{} {{Part}} 22: {{Cognitive Wireless RAN Medium Access
  Control}} ({{MAC}}) and {{Physical Layer}} ({{PHY}}) specifications:
  {{Policies}} and procedures for operation in the {{TV Bands}},'' \emph{IEEE
  Std 80222-2011}, pp. 1--680, Jul. 2011.

\bibitem{fcc_cbrs_2015}
{Federal Communications Commission}, ``Amendment of the {{Commission}}'s
  {{Rules}} with {{Regard}} to {{Commercial Operations}} in the 3550-3650 {{MHz
  Band}},'' Federal Communications Commission, REPORT AND ORDER AND SECOND
  FURTHER NOTICE OF PROPOSED RULEMAKING 15-47, Apr. 2015.

\bibitem{laghate_cooperatively_2017}
M.~Laghate and D.~Cabric, ``Cooperatively {{Learning Footprints}} of {{Multiple
  Incumbent Transmitters}} by {{Using Cognitive Radio Networks}},'' \emph{IEEE
  Trans. Cogn. Commun. Netw.}, vol.~PP, no.~99, pp. 1--1, 2017.

\bibitem{tilghman_inferring_2013}
P.~Tilghman and D.~Rosenbluth, ``Inferring {{Wireless Communications Links}}
  and {{Network Topology}} from {{Externals Using Granger Causality}},'' in
  \emph{{{IEEE MILCOM}}}, Nov. 2013, pp. 1284--1289.

\bibitem{moore_analysis_2016}
M.~G. Moore and M.~A. Davenport, ``Analysis of wireless networks using
  {{Hawkes}} processes,'' in \emph{{{IEEE SPAWC}}}, Jul. 2016, pp. 1--5.

\bibitem{kokalj-filipovic_learning_2016}
S.~Kokalj-Filipovic, C.~B. Acosta, and M.~Pepe, ``Learning structural
  properties of wireless ad-hoc networks non-parametrically from spectral
  activity samples,'' in \emph{{{IEEE GlobalSIP}}}, Dec. 2016, pp. 1092--1097.

\bibitem{ieee_80211_2016}
``{{IEEE Standard}} for {{Information}} technology--{{Telecommunications}} and
  information exchange between systems {{Local}} and metropolitan area
  networks--{{Specific}} requirements - {{Part}} 11: {{Wireless LAN Medium
  Access Control}} ({{MAC}}) and {{Physical Layer}} ({{PHY}})
  {{Specifications}},'' \emph{IEEE Std 80211-2016 Revis. IEEE Std 80211-2012},
  pp. 1--3534, Dec. 2016.

\bibitem{etsi_lte_2017}
``{{LTE}}; {{Evolved Universal Terrestrial Radio Access}} ({{E}}-{{UTRA}});
  {{Physical}} channels and modulation ({{3GPP TS}} 36.211 version 14.2.0
  {{Release}} 14),'' European Telecommunications Standards Institute, Tech.
  Rep. RTS/TSGR-0136211ve20, Apr. 2017.

\bibitem{bluetoothsig_bluetooth_2016}
{Bluetooth SIG}, ``Bluetooth {{Core Specification}} v5.0,'' Dec. 2016.

\bibitem{granger_investigating_1969}
C.~W.~J. Granger, ``Investigating {{Causal Relations}} by {{Econometric
  Models}} and {{Cross}}-spectral {{Methods}},'' \emph{Econometrica}, vol.~37,
  no.~3, pp. 424--438, 1969.

\bibitem{sayed_adaptive_2011}
A.~Sayed, \emph{Adaptive {{Filters}}}.\hskip 1em plus 0.5em minus 0.4em\relax
  {Wiley}, 2011.

\bibitem{seth_matlab_2010}
A.~K. Seth, ``A {{MATLAB}} toolbox for {{Granger}} causal connectivity
  analysis,'' \emph{J. Neurosci. Methods}, vol. 186, no.~2, pp. 262--273, 2010.

\bibitem{geweke_measurement_1982}
J.~Geweke, ``Measurement of {{Linear Dependence}} and {{Feedback}} between
  {{Multiple Time Series}},'' \emph{J. Am. Stat. Assoc.}, vol.~77, no. 378, pp.
  304--313, 1982.

\bibitem{eichler_graphical_2017}
M.~Eichler, R.~Dahlhaus, and J.~Dueck, ``Graphical {{Modeling}} for
  {{Multivariate Hawkes Processes}} with {{Nonparametric Link Functions}},''
  \emph{J. Time Ser. Anal.}, vol.~38, no.~2, pp. 225--242, 2017.

\bibitem{xu_learning_2016}
H.~Xu, M.~Farajtabar, and H.~Zha, ``Learning {{Granger Causality}} for {{Hawkes
  Processes}},'' in \emph{{{ICML}}}, ser. ICML'16, vol.~48.\hskip 1em plus
  0.5em minus 0.4em\relax New York, NY, USA: {JMLR.org}, 2016, pp. 1660--1669.

\bibitem{schreiber_measuring_2000}
T.~Schreiber, ``Measuring {{Information Transfer}},'' \emph{Phys Rev Lett},
  vol.~85, no.~2, pp. 461--464, Jul. 2000.

\bibitem{hatemi-j_asymmetric_2012}
A.~Hatemi-J, ``Asymmetric causality tests with an application,'' \emph{Empir.
  Econ.}, vol.~43, no.~1, pp. 447--456, 2012.

\bibitem{barnett_transfer_2012}
L.~Barnett and T.~Bossomaier, ``Transfer {{Entropy}} as a {{Log}}-{{Likelihood
  Ratio}},'' \emph{Phys Rev Lett}, vol. 109, no.~13, p. 138105, Sep. 2012.

\bibitem{wald_tests_1943}
A.~Wald, ``Tests of statistical hypotheses concerning several parameters when
  the number of observations is large,'' \emph{Trans. Amer. Math. Soc.},
  vol.~54, pp. 426--482, 1943.

\end{thebibliography}
\end{document}